\documentclass[a4paper,preprint,superscriptaddress,nofootinbib,pra,11pt]{revtex4-1}

\usepackage[dvipdfmx]{graphicx} 
\usepackage{bm}
\usepackage{amssymb}
\usepackage{amsmath}
\usepackage{caption}
\usepackage{color}
\usepackage{cancel}
\usepackage{comment}
\usepackage{here}
\usepackage{multirow}
\usepackage{url}
\usepackage{braket}
\usepackage{tikz}
\usepackage{float} 
\usepackage{tikz-feynhand}
\usepackage{tikz-feynman}
\usepackage[subrefformat=parens]{subcaption}
\usepackage{mathrsfs}
\usepackage[colorlinks=true,allcolors=blue]{hyperref}
\usepackage{ulem}

\usepackage[toc,page]{appendix}

\allowdisplaybreaks

\begin{document} 

\title{Loop induced $H^\pm W^\mp Z$ vertices in the general two Higgs doublet model with CP violation}

\author{Shinya Kanemura}
\email{kanemu@het.phys.sci.osaka-u.ac.jp }
\affiliation{Department of Physics, Osaka University, Toyonaka, Osaka 560-0043, Japan}

\author{Yushi Mura}
\email{y\_mura@het.phys.sci.osaka-u.ac.jp }
\affiliation{Department of Physics, Osaka University, Toyonaka, Osaka 560-0043, Japan}

\preprint{OU-HET-1241}
 
\begin{abstract}
Understanding symmetry of extended Higgs models helps to construct the theory beyond the standard model and gives important insights for testing these models by experiments.
The custodial symmetry plays an important role as a leading principle to construct non-minimal Higgs sectors.
On the other hand, new sources of CP violation in the Higgs sector are essentially important to explain the baryon asymmetry of the Universe.
In the two Higgs doublet model, it has been known that the CP violation and the custodial symmetry are not compatible, and thus the CP violating Higgs potential in general violates the custodial symmetry in the Higgs sector.
In this paper, we discuss the loop-induced $H^\pm W^\mp Z$ vertices in the two Higgs doublet model, which are induced due to the custodial symmetry violation of the sector of the particles in the loop, and study how the vertices are affected by the CP violation.
We calculate the $H^\pm W^\mp Z$ vertices in the most general two Higgs doublet model with the CP violation at the one-loop level and evaluate the decay modes $H^\pm \to W^\pm Z$ caused by these vertices. 
We obtain new contributions to the $H^\pm W^\mp Z$ vertices from the CP violating part of the model, in addition to the known contributions from the CP conserving part.
Moreover, we find that an asymmetry between the decays $H^+ \to W^+ Z$ and $H^- \to W^- Z$ is caused by the CP violating phases in the model, some of which are important for the electroweak baryogenesis.
We also briefly mention testability of these vertices at future collider experiments.
\end{abstract}

\maketitle

\section{Introduction\label{sec:intro}}
The Standard Model (SM) was established by the discovery of the Higgs boson at the LHC in 2012~\cite{ATLAS:2012yve,CMS:2012qbp}.
There are still some unsolved problems in particle physics and cosmology.
Models of new physics beyond the SM are needed to solve these problems.
For such a case, discussions on symmetries usually give important insights into understanding physics.
Extended Higgs sectors have often been considered in these new physics models, and the way of extension can be classified by global symmetries in the Higgs potential.
For example, the custodial symmetry~\cite{Sikivie:1980hm}, which is a global symmetry remaining after the electroweak symmetry breaking, and its violation are important for phenomenology to know the Higgs sector.

In the SM, the Higgs potential has the custodial symmetry, and consequently, the masses of the electroweak gauge bosons are related: i.e. $\rho \equiv m_W^2 / m_Z^2 \cos^2 \theta_W = 1$ at the tree level, where $m_W$ and $m_Z$ are the masses of the W and Z bosons, respectively, and $\theta_W$ is the Weinberg angle which measures violation of the custodial symmetry due to the hypercharge.
It is well known that this relation is relaxed by the radiative corrections~\cite{Veltman:1977kh,Chanowitz:1978uj,Toussaint:1978zm,Lytel:1980zh,Einhorn:1981cy,Peskin:1990zt,Peskin:1991sw}.
By the dedicated experimental efforts for the $\rho$ parameter, $\rho \simeq 1$ has been observed~\cite{Baak:2012kk,Baak:2014ora,ParticleDataGroup:2022pth}, so that any models for new physics have to respect this fact.
As one of the extended Higgs models, the two Higgs Doublet Model (2HDM) is often considered.
The custodial symmetry in the 2HDM has been discussed in refs.~\cite{Haber:1992py,Pomarol:1993mu,Haber:2010bw,Gerard:2007kn,Grzadkowski:2010dj,Kanemura:2011sj,Aiko:2020atr}, and the $\rho$ parameter is deviated from the SM prediction due to the loop effects from additional scalar bosons~\cite{Lytel:1980zh,Toussaint:1978zm,Pomarol:1993mu,Haber:2010bw}.

The 2HDM may be motivated by solving the remaining problems, such as the Baryon Asymmetry of the Universe (BAU).
Baryogenesis is one of the promising scenarios for the BAU.
In the scenario of baryogenesis, CP violation is required by the Sakharov's third conditions~\cite{Sakharov:1967dj}.
However, the CP violating phase in the quark sector in the SM is too small to explain the observed baryon asymmetry~\cite{Huet:1994jb} with the mechanism of electroweak baryogenesis~\cite{Kuzmin:1985mm}.
In the 2HDM, on the other hand, new sources of the CP violation are naturally introduced, so that various electroweak baryogenesis models have been discussed in the literature~\cite{Turok:1990zg,Cline:1995dg,Fromme:2006cm,Cline:2011mm,Tulin:2011wi,Liu:2011jh,Ahmadvand:2013sna,Chiang:2016vgf,Guo:2016ixx,Fuyuto:2017ewj,Dorsch:2016nrg,Modak:2018csw,Basler:2021kgq,Enomoto:2021dkl,Enomoto:2022rrl,Kanemura:2023juv,Aoki:2023xnn}.

It has been known that the CP violating potential in the 2HDM violates the custodial symmetry~\cite{Pomarol:1993mu}.
Therefore, in the CP violating 2HDM, some physical observables are induced at the loop level due to the custodial symmetry violation, e.g., corrections to the $\rho$ parameter~\cite{Pomarol:1993mu}, the $H^\pm W^\mp Z$ vertices~\cite{Grifols:1980uq}, and so on.
Such an observable can be an important tool to understand the structure of the Higgs sector.

In some classes of non-minimal Higgs sectors, the existence of $H^\pm$ is phenomenologically important.  
The $H^\pm W^\mp Z$ vertices have been well studied as a clear observable for the violation of the custodial symmetry.
In the 2HDM, these vertices are absent at the tree level~\cite{Grifols:1980uq}.
However, it can be induced by loop effects~\cite{Rizzo:1989ym}.
These $H^\pm W^\mp Z$ vertices have been analyzed in the context of the CP conserving 2HDM~\cite{Rizzo:1989ym,CapdequiPeyranere:1990qk,Kanemura:1997ej,Diaz-Cruz:2001thx,Arhrib:2006wd,Abbas:2018pfp,Aiko:2021can}, the Minimal Supersymmetric Standard Model (MSSM)~\cite{Mendez:1990epa,CapdequiPeyranere:1990qk,Arhrib:2007ed}, and the CP violating MSSM~\cite{Arhrib:2006wd,Arhrib:2007rm}.
In general, such a one-loop-induced vertex is suppressed by the factor $1/16\pi^2$, so that one might expect that effects from these vertices are small.
In the 2HDM, however, it has been pointed out that these vertices can be enhanced by non-decoupling quantum effects of particles in the loop~\cite{Kanemura:1997ej}.
Besides the 2HDM, the feature of the $H^\pm W^\mp Z$ vertices have been studied in the models with $\mathrm{SU}(2)_L$ triplets~\cite{Grifols:1980uq,Mohapatra:1979ia,Mohapatra:1980yp,Georgi:1985nv,Chanowitz:1985ug}, the three Higgs doublet model~\cite{Moretti:2015tva}, and the model with one colored scalar boson~\cite{Chakrabarty:2020msl}.
Through these vertices, at high energy colliders, the charged scalar bosons can be produced, and they can decay into the W and Z bosons.
Phenomenological impacts of these vertices at the collider experiments have been discussed in refs.~~\cite{Godbole:1994np,Cheung:1994rp,Ghosh:1996jg,Kanemura:1999tg,Kanemura:2000cw,Huitu:2000ut,Cheung:2002gd,Logan:2002jh,Asakawa:2005gv,Asakawa:2006gm,Kanemura:2011kc,Chiang:2014bia,Chiang:2015rva,Adhikary:2020cli}.
From the current ATLAS~\cite{ATLAS:2022zuc} and CMS~\cite{CMS:2021wlt} data related to the $H^\pm W^\mp Z$ vertices, some of the models, e.g., the Georgi--Machacek model~\cite{Georgi:1985nv,Chanowitz:1985ug}, have been already constrained.

In this paper, we discuss the $H^\pm W^\mp Z$ vertices in the most general 2HDM with the CP violation, in which the custodial symmetry is also violated by the CP violating effect.
Such a CP violating effect can be a new source of the $H^\pm W^\mp Z$ vertices.
We calculate the $H^\pm W^\mp Z$ vertices at the one-loop level, and we confirm that there are both contributions from the CP conserving and CP violating coupling constants.
We then discuss the decays $H^\pm \to W^\pm Z$.
We find that these decays are significantly enhanced by the loop effects of the additional scalar bosons.
Furthermore, we find that an asymmetry between the decay rates of $H^+ \to W^+ Z$ and $H^- \to W^- Z$ can be a useful observable to see the CP violating phases in the model. 
Finally, we give a short discussion for testability of these vertices at current and future high energy collider experiments.

This paper is organized as follows.
In section~\ref{sec:model}, we define the most general 2HDM with the CP violation.
In section~\ref{sec:symmetry}, we introduce the custodial symmetry and the CP symmetry in the model.
Subsequently, physical observables for the custodial symmetry violation, including the $H^\pm W^\mp Z$ vertices, are introduced.
We discuss the relation among those symmetry and observables in this section.
In section~\ref{sec:vertex}, we show the calculation of the $H^\pm W^\mp Z$ vertices in our model, and show numerical results for the decays $H^\pm \to W^\pm Z$.
We also present our results for the CP asymmetry between the decay rates of $H^+ \to W^+ Z$ and $H^- \to W^- Z$.
In section~\ref{sec:discussion}, we give some discussions and comments on the results obtained in the previous sections.
The definition of the loop functions and the full formulae for the $H^\pm W^\pm Z$ vertices are shown in appendices~\ref{sec:app2} and~\ref{sec:app1}, respectively.

\section{The model \label{sec:model}}

We here define the most general two Higgs doublet model (2HDM), in which two $\mathrm{SU}(2)_L$ doublets $\Phi_k = (\phi_k^+, \phi_k^0)^T~(k=1,2)$ with the hypercharge $Y=1/2$ are introduced.
After the spontaneous electroweak symmetry breaking, without loss of generality, we can take the Higgs basis~\cite{Davidson:2005cw} as 
\begin{align}
  \Phi_1 = 
  \begin{pmatrix}
     G^+ \\
     \frac{1}{\sqrt{2}} ( v + h_1 + i G^0)
  \end{pmatrix},
  ~~
  \Phi_2 = 
  \begin{pmatrix}
     H^+ \\
     \frac{1}{\sqrt{2}} ( h_2 + i h_3)
  \end{pmatrix},
\end{align}
where $v \simeq 246$ GeV is the Vacuum Expectation Value (VEV), $G^+$, $G^0$ are the Nambu--Goldstone bosons (NG bosons), $H^\pm$ are the charged scalar bosons, and $h_i~(i=1,2,3)$ are the electric neutral scalar bosons.
In the Higgs basis, the most general Higgs potential is given by 
\begin{align}
  V = &-Y_1^2 (\Phi_1^\dagger \Phi_1) - Y_2^2 (\Phi_2^\dagger \Phi_2) - \Big(Y_3^2 (\Phi_1^\dagger \Phi_2) + \mathrm{h.c.} \Big) \notag \\
  &+\frac{1}{2}Z_1 (\Phi_1^\dagger \Phi_1)^2 +\frac{1}{2}Z_2 (\Phi_2^\dagger \Phi_2)^2 + Z_3 (\Phi_1^\dagger \Phi_1)(\Phi_2^\dagger \Phi_2) + Z_4 (\Phi_2^\dagger \Phi_1)(\Phi_1^\dagger \Phi_2) \notag \\
  &+ \bigg\{ \Big( \frac{1}{2}Z_5 \Phi_1^\dagger \Phi_2 + Z_6 \Phi_1^\dagger \Phi_1 + Z_7 \Phi_2^\dagger \Phi_2 \Big) \Phi_1^\dagger \Phi_2 + \mathrm{h.c.} \bigg\},
  \label{eq:potential}
\end{align}
where $Y_1^2 (>0)$, $Y_2^2$, $Z_1$, $Z_2$, $Z_3$ and $Z_4$ are real parameters, and $Y_3^2$, $Z_5$, $Z_6$ and $Z_7$ are complex parameters.
By using a rotation of the phase for $\Phi_2$, we can make $Z_5$ real.
In the following, we call this basis of $\Phi_2$ as \textit{the real $Z_5$ basis}.
From the stationary condition, we obtain
\begin{align}
  Y_1^2 = \frac{1}{2} Z_1 v^2, ~~~~ Y_3^2 = \frac{1}{2} Z_6 v^2.
\end{align}

The squared mass of $H^\pm$ is given by
\begin{align}
  m_{H^\pm}^2 = -Y_2^2 + \frac{1}{2} Z_3 v^2,
  \label{eq:chargedmass}
\end{align}
and the squared mass matrix for $h_i$ in the real $Z_5$ basis is given by 
\begin{align}
  \mathcal{M}_{ij}^2 \equiv \frac{\partial^2 V}{\partial h_i \partial h_j} &= 
  \begin{pmatrix}
      Z_1 v^2 & Z_6^R v^2 & - Z_6^I v^2 \\
      Z_6^R v^2 & m_{H^\pm}^2 +\frac{1}{2}(Z_4 + Z_5) v^2 & 0 \\
      - Z_6^I v^2 & 0 & m_{H^\pm}^2 +\frac{1}{2}(Z_4 - Z_5) v^2
  \end{pmatrix},
  \label{eq:neut_mass}
\end{align}
where the subscripts $R$ ($I$) mean the real (imaginary) part of the coupling constants.
By using an orthogonal matrix $\mathcal{R}$ which represents $\mathrm{SO}(3)$ transformation, the mass eigenstates for the neutral scalar bosons $H_i$ are defined as 
\begin{align}
  H_i = \mathcal{R}_{ij} h_j.
  \label{eq:definemasseigen}
\end{align}
The matrix $\mathcal{R}$ can be parametrized by 
\begin{align}
  \mathcal{R} &= \mathcal{R}_1 \mathcal{R}_2 \mathcal{R}_3 \notag \\
  &\equiv 
  \begin{pmatrix}
      \cos \alpha_1 & -\sin \alpha_1 & 0 \\
      \sin \alpha_1 & \cos \alpha_1 & 0 \\
      0 & 0 & 1
  \end{pmatrix}
  \begin{pmatrix}
      \cos \alpha_2 & 0 & -\sin \alpha_2 \\
      0 & 1 & 0 \\
      \sin \alpha_2 & 0 & \cos \alpha_2
  \end{pmatrix}
  \begin{pmatrix}
      1 & 0 & 0 \\
      0 & \cos \alpha_3 & -\sin \alpha_3 \\
      0 & \sin \alpha_3 & \cos \alpha_3
  \end{pmatrix} \notag \\
  &=
  \begin{pmatrix}
      \cos \alpha_2 \cos \alpha_1 & - \cos \alpha_3 \sin \alpha_1 - \cos \alpha_1 \sin \alpha_2 \sin \alpha_3 & -\cos \alpha_1 \cos \alpha_3 \sin \alpha_2 + \sin \alpha_1 \sin \alpha_3 \\
      \cos \alpha_2 \sin \alpha_1 & \cos \alpha_1 \cos \alpha_3 - \sin \alpha_1 \sin \alpha_2 \sin \alpha_3 & -\cos \alpha_3 \sin \alpha_1 \sin \alpha_2 - \cos \alpha_1 \sin \alpha_3 \\
      \sin \alpha_2 & \cos \alpha_2 \sin \alpha_3 & \cos \alpha_2 \cos \alpha_3
  \end{pmatrix},
  \label{eq:Rmatrix}
\end{align}
where $-\pi \le \alpha_1, \alpha_3 < \pi$, $-\pi/2 \le \alpha_2 < \pi/2$~\cite{Haber:2006ue}, and it satisfies 
\begin{align}
  \mathcal{R} \mathcal{M}^2 \mathcal{R}^T = \mathrm{diag}(m_{H_1}^2, m_{H_2}^2, m_{H_3}^2).
  \label{eq:relation_neutmass_diagmass}
\end{align}
We identify $H_1$ with the discovered Higgs boson at the LHC~\cite{ATLAS:2012yve,CMS:2012qbp}, so that we set $m_{H_1} = 125$~GeV.

In the Higgs basis, the most general Yukawa interaction is given by 
\begin{align}
  -\mathcal{L}_Y = \sum_{k=1}^2 \left( \overline{Q^\prime_{L}} Y^{u\dagger}_{k} \tilde{\Phi}_k u^\prime_{R} + \overline{Q^\prime_{L}} Y^{d}_{k} \Phi_k d^\prime_{R} + \overline{L^\prime_{L}} Y^e_{k} \Phi_k e^\prime_{R} + \mathrm{h.c.} \right),
\end{align}
where $\tilde{\Phi}_k \equiv i \sigma_2 \Phi^*_k = \big((\phi^0_k)^*, -\phi^-_k \big)$, and $Q_L^\prime$ and $L_L^\prime$ ($u_R^\prime$, $d_R^\prime$ and $e_R^\prime$) are the left- (right-) handed quarks and the leptons in the gauge eigenstates, respectively.
We here have omitted the fermion-flavor indices.
The matrices $Y_1^f$ and $Y_2^f$ ($f=u,d,e$) are the $3\times3$ Yukawa matrices.
In the mass eigenstate of the fermions, in which $Y_1^f$ are diagonalized, the Yukawa interaction is written by 
\begin{align}
  -\mathcal{L}_Y &=  \overline{Q_{L}^u} Y^u_{\mathrm{d}} \tilde{\Phi}_1 u_{R} + \overline{Q_{L}^d} Y^{d}_{\mathrm{d}} \Phi_1 d_{R} + \overline{L_{L}} Y^e_{\mathrm{d}} \Phi_1 e_{R} + \mathrm{h.c.}  \notag \\
  &+ \overline{Q_{L}^u} \rho^u \tilde{\Phi}_2 u_{R} + \overline{Q_{L}^d} \rho^{d} \Phi_2 d_{R} + \overline{L_{L}} \rho^e \Phi_2 e_{R} + \mathrm{h.c.},
\end{align}
where the $\mathrm{SU}(2)_L$ doublets are defined by $Q_L^u =  (u_L, V d_L)^T$, $Q_L^d =  ( V^\dagger u_L,d_L)^T$ and $L_L = (\nu_L, e_L)^T$ with the Cabibbo--Kobayashi--Masukawa matrix $V$~\cite{Cabibbo:1963yz,Kobayashi:1973fv}.
The diagonalized matrices $Y_{\mathrm{d}}^f$ is written by $Y_{\mathrm{d}}^f = \frac{\sqrt{2}}{v} \mathrm{diag}(m_{f_1}, m_{f_2}, m_{f_3})$.
On the other hand, $\rho^f$ are not diagonalized in this mass eigenstate.
We parametrize the $\rho^f$ matrices as 
\begin{align}
  \rho^u = 
  \begin{pmatrix}
  \rho_{uu} & \rho_{uc} & \rho_{ut} \\
  \rho_{cu} & \rho_{cc} & \rho_{ct} \\
  \rho_{tu} & \rho_{tc} & \rho_{tt}
  \end{pmatrix}, 
  ~~
  \rho^d = 
  \begin{pmatrix}
  \rho_{dd} & \rho_{ds} & \rho_{db} \\
  \rho_{sd} & \rho_{ss} & \rho_{sb} \\
  \rho_{bd} & \rho_{bs} & \rho_{bb}
  \end{pmatrix},
  ~~
  \rho^e = 
  \begin{pmatrix}
  \rho_{ee} & \rho_{e\mu} & \rho_{e\tau} \\
  \rho_{\mu e} & \rho_{\mu \mu} & \rho_{\mu \tau} \\
  \rho_{\tau e} & \rho_{\tau \mu} & \rho_{\tau \tau}
  \end{pmatrix},
\end{align}
and the components of each matrix are generally complex.

\begin{table}[t]
  \centering
  \begin{tabular}{|c|c|c|c|c|}
      \hline
       & $\big(\rho^u \big)_{ll}$ & $\big(\rho^d \big)_{ll}$ & $\big(\rho^e \big)_{ll}$ & $\big(\rho^f \big)_{lm} (l \neq m)$  \\ \hline
  Type-I &  $\big( Y_{\mathrm{d}}^u \big)_{ll} \cot \beta$    &  $\big( Y_{\mathrm{d}}^d \big)_{ll} \cot \beta$  &  $\big( Y_{\mathrm{d}}^e \big)_{ll} \cot \beta$  &   0    \\ \hline
  Type-II &  $\big( Y_{\mathrm{d}}^u \big)_{ll} \cot \beta$    &  $-\big( Y_{\mathrm{d}}^d \big)_{ll} \tan \beta$   & $-\big( Y_{\mathrm{d}}^e \big)_{ll} \tan \beta$  & 0  \\ \hline
  Type-X & $\big( Y_{\mathrm{d}}^u \big)_{ll} \cot \beta$ & $\big( Y_{\mathrm{d}}^d \big)_{ll} \cot \beta$ & $-\big( Y_{\mathrm{d}}^e \big)_{ll} \tan \beta$ & 0 \\ \hline 
  Type-Y & $\big( Y_{\mathrm{d}}^u \big)_{ll} \cot \beta$ & $-\big( Y_{\mathrm{d}}^d \big)_{ll} \tan \beta$ & $\big( Y_{\mathrm{d}}^e \big)_{ll} \cot \beta$ &  0 \\ \hline
  \end{tabular}
  \caption{The correspondence of $\big( \rho^f \big)_{lm}$ ($f =  u,d,e$) in our model to the ones in the softly-broken $\mathbb{Z}_2$ symmetric 2HDM (Type-I, II, X, and Y).}
  \label{tab:Z2}
\end{table} 
The off-diagonal elements of these matrices cause the flavor changing neutral current, and it is severely constrained by the flavor experiments~\cite{Crivellin:2013wna}.
As one of the way to avoid this difficulty, the softly-broken $\mathbb{Z}_2$ symmetry in the 2HDM is often considered~\cite{Lee:1973iz,Glashow:1976nt,Paschos:1976ay,Haber:1978jt,Donoghue:1978cj,Hall:1981bc,Gunion:1989we,Barger:1989fj,Aoki:2009ha,Branco:2011iw}.
There are four types (Type-I, II, X, and Y~\cite{Aoki:2009ha}) of the assignment of the $\mathbb{Z}_2$ parity.
As summarized in table~\ref{tab:Z2}, the additional Yukawa matrices $\rho^f$ can be written by using $Y^f_{\mathrm{d}}$ and the mixing angle $\beta$ from the $\mathbb{Z}_2$ basis to the Higgs basis.
In our model, due to the absence of the softly-broken $\mathbb{Z}_2$ symmetry, the CP violating phases can be introduced in the $\rho^f$ matrices.
However, the off-diagonal elements must be small to satisfy the experimental data, which are discussed in the later section.

The kinetic term of $\Phi_1$ and $\Phi_2$ is given by 
\begin{align}
  \mathcal{L}_k = |D_\mu \Phi_1|^2 + |D_\mu \Phi_2|^2.
\end{align}
The covariant derivative is defined as
\begin{align}
  D_\mu = \partial_\mu - i g \frac{\sigma^a}{2} W_\mu^a - i g^\prime \frac{1}{2} B_\mu,
\end{align}
where $\sigma^a$ ($a=1,2,3$) is the Pauli matrices, and $g$ and $g^\prime$ are the $\mathrm{SU}(2)_L$ and $\mathrm{U}(1)_Y$ gauge coupling constants, respectively.

The coupling constants for $H_1$ have been measured at the LHC.
The ATLAS and CMS Run~2 data~\cite{ATLAS:2022vkf,CMS:2022dwd} are consistent with the SM value.
In our model, the coupling constants $H_1 VV$ ($V=W^\pm, Z$) are modified by the factor $\mathcal{R}_{11}$ from the SM predictions, as indicated in eqs.~(\ref{eq:definemasseigen}) and (\ref{eq:Rmatrix}).
The $H_1 ff$ coupling constants are also modified from the SM, and the Yukawa interactions for $H_1$ are given by
\begin{align}
  \mathcal{L}_Y \supset - \frac{1}{\sqrt{2}} \Big( \mathcal{R}_{11} Y_{\mathrm{d}}^f + (\mathcal{R}_{12} \mp i\mathcal{R}_{13} ) \rho^f  \Big) H_1 \overline{f}_L f_R + \mathrm{h.c.},
\end{align}
where the upper (lower) sign is the case for the up-type quarks (the down-type quarks and the leptons).

\section{The custodial symmetry and the loop-induced $H^\pm W^\mp Z$ vertices \label{sec:symmetry}}

In this section, we first review the custodial symmetry and the CP symmetry in the general 2HDM.
We discuss the relation between these symmetries, and show that the CP violating potential violates the custodial symmetry.
We then introduce some physical observables, e.g. $\Delta \rho$ ($= \rho - 1$) and the $H^\pm W^\mp Z$ vertices, as the consequences of the violation of the custodial symmetry.
Previously, the loop-induced $H^\pm W^\mp Z$ vertices and the decays $H^\pm \to W^\pm Z$ have been analyzed in the CP conserving 2HDM~\cite{Rizzo:1989ym,CapdequiPeyranere:1990qk,Kanemura:1997ej,Diaz-Cruz:2001thx,Arhrib:2006wd,Abbas:2018pfp,Aiko:2021can}.
At the end of this section, we discuss the $H^\pm W^\mp Z$ vertices caused by the CP violating potential and show relations among the custodial symmetry violation, the CP violation, and the $H^\pm W^\mp Z$ vertices.

\subsection{The conditions for the custodial and CP symmetry}

According to ref.~\cite{Haber:2010bw}, we derive the conditions for the custodial symmetry of the potential.
We introduce bilinear forms as
\begin{align}
  \mathbb{M}_1 \equiv \big(\tilde{\Phi}_1, \Phi_1 \big), ~~  \mathbb{M}_2 \equiv \big(\tilde{\Phi}_2, \Phi_2 \big)
  \begin{pmatrix}
    e^{- i\chi}  & 0 \\
    0 & e^{i\chi}
  \end{pmatrix},
\end{align}
where the scalar doublets $\Phi_1$ and $\Phi_2$ are defined in the Higgs basis (cf. eq.~(\ref{eq:potential})).
The phase $\chi$ ($0 \le \chi < 2\pi$) represents the degree of freedom of the phase rotation for $\Phi_2$.
The transformation law of global $\mathrm{SU}(2)_L \times \mathrm{SU}(2)_R$ for $\mathbb{M}_1$ and $\mathbb{M}_2$ is $\mathbb{M}_{1,2} \to L \mathbb{M}_{1,2} R^\dagger$, where $L \in \mathrm{SU}(2)_L$ and $R \in \mathrm{SU}(2)_R$.
After the spontaneous electroweak symmetry breaking, these matrices acquire the VEV as $\langle \mathbb{M}_1 \rangle = (v/\sqrt{2}) \times I_2$ and $\langle \mathbb{M}_2 \rangle = 0$, where $I_2$ is the $2 \times 2$ identical matrix.
As a result, these matrices are invariant under the $L=R$ transformation, i.e. the custodial $\mathrm{SU}(2)_V$ transformation, even after the electroweak symmetry breaking.
We can construct $\mathrm{SU}(2)_L \times \mathrm{U}(1)_Y$ gauge invariants as
\begin{align}
  &\mathrm{Tr}[\mathbb{M}_1^\dagger \mathbb{M}_1] = 2 |\Phi_1|^2, ~~~~ \mathrm{Tr}[\mathbb{M}_2^\dagger \mathbb{M}_2] = 2 |\Phi_2|^2, ~~~~\mathrm{Tr}[\mathbb{M}_1^\dagger \mathbb{M}_2] = e^{i \chi} \Phi_1^\dagger \Phi_2 + e^{-i\chi} \Phi_2^\dagger \Phi_1, \notag \\
  &\mathrm{Tr}[\mathbb{M}_1^\dagger \mathbb{M}_2 \sigma_3] = e^{i \chi} \Phi_1^\dagger \Phi_2 - e^{-i\chi} \Phi_2^\dagger \Phi_1.
  \label{eq:invariants}
\end{align}
The invariants in the first line of eq.~(\ref{eq:invariants}) are custodial $\mathrm{SU}(2)_V$ invariants, but in the second line are not.
By using them, the potential which is given in eq.~(\ref{eq:potential}) can be rewritten as 
\begin{align}
  V = &-\frac{1}{2} Y_1^2 \mathrm{Tr}[\mathbb{M}_1^\dagger \mathbb{M}_1] -\frac{1}{2} Y_2^2 \mathrm{Tr}[\mathbb{M}_2^{\dagger} \mathbb{M}_2] - \mathrm{Re}[Y_3^2 e^{-i\chi}] \mathrm{Tr}[\mathbb{M}_1^\dagger \mathbb{M}_2] \notag \\
  &+\frac{1}{8}Z_1 \mathrm{Tr}[\mathbb{M}_1^\dagger \mathbb{M}_1]^2 +\frac{1}{8}Z_2 \mathrm{Tr}[\mathbb{M}_2^{\dagger} \mathbb{M}_2]^2 + \frac{1}{4} Z_3 \mathrm{Tr}[\mathbb{M}_1^\dagger \mathbb{M}_1]\mathrm{Tr}[\mathbb{M}_2^{\dagger} \mathbb{M}_2] \notag \\
  &+ \frac{1}{4} (Z_4 + \mathrm{Re}[Z_5 e^{-2i \chi}]) \mathrm{Tr}[\mathbb{M}_1^\dagger \mathbb{M}_2]^2  \notag \\
  &+\frac{1}{2} \Big( \mathrm{Re}[Z_6 e^{- i \chi}] \mathrm{Tr}[\mathbb{M}_1^\dagger \mathbb{M}_1] +\mathrm{Re}[Z_7 e^{- i \chi}] \mathrm{Tr}[\mathbb{M}_2^{\dagger} \mathbb{M}_2] \Big) \mathrm{Tr}[\mathbb{M}_1^\dagger \mathbb{M}_2] \notag \\
  &+i \mathrm{Im}[Y_3^2 e^{- i \chi}] \mathrm{Tr}[\mathbb{M}_1^\dagger \mathbb{M}_2 \sigma_3] - \frac{1}{4} (Z_4 - \mathrm{Re}[Z_5 e^{-2i \chi}]) \mathrm{Tr}[\mathbb{M}_1^\dagger \mathbb{M}_2 \sigma_3]^2 \notag \\
  &+\frac{i}{2} \mathrm{Im}[Z_5 e^{-2 i \chi}] \mathrm{Tr}[\mathbb{M}_1^\dagger \mathbb{M}_2] \mathrm{Tr}[\mathbb{M}_1^\dagger \mathbb{M}_2 \sigma_3] \notag \\
  &+\frac{i}{2} \Big( \mathrm{Im}[Z_6 e^{- i \chi}] \mathrm{Tr}[\mathbb{M}_1^\dagger \mathbb{M}_1] + \mathrm{Im}[Z_7 e^{- i \chi}] \mathrm{Tr}[\mathbb{M}_2^{\dagger} \mathbb{M}_2] \Big) \mathrm{Tr}[\mathbb{M}_1^\dagger \mathbb{M}_2 \sigma_3].
  \label{eq:custodial_potential}
\end{align}
Obviously, the last three lines in eq.~(\ref{eq:custodial_potential}) are not custodial $\mathrm{SU}(2)_V$ invariant, so that we obtain the conditions for the custodial symmetric potential:
\begin{align}
  &\mathrm{Im}[Y_3^2 e^{- i \chi}] = \mathrm{Im}[Z_5 e^{-2 i \chi}] =  \mathrm{Im}[Z_6 e^{- i \chi}] = \mathrm{Im}[Z_7 e^{- i \chi}] = 0, \notag \\
  &Z_4 = \mathrm{Re}[Z_5 e^{-2 i \chi}].
\end{align}
In the real $Z_5$ basis, the conditions are 
\begin{align}
  &\mathrm{Custodial~symmetry}: \hspace{1.65cm} Z_4 = Z_5, ~~~~ Z_6^I = Z_7^I = 0 ~~~~ (\chi = 0, ~\pi),
  \label{eq:condition_custodial} \\
  &\mathrm{Twisted~custodial~symmetry}:~~Z_4 = -Z_5, ~~ Z_6^R = Z_7^R = 0 ~~~ (\chi = \pi/2, ~3 \pi/2).
  \label{eq:condition_twist_custodial}
\end{align}
The conditions in the eq.~(\ref{eq:condition_custodial}) (eq.~(\ref{eq:condition_twist_custodial})) for $\chi = 0$ and $\pi$ ($\chi = \pi/2$ and $3\pi/2$) have been known as the conditions for the (twisted) custodial symmetry in the potential~\cite{Pomarol:1993mu,Gerard:2007kn,Haber:2010bw}.
We note, even if the potential satisfies eq.~(\ref{eq:condition_custodial}) or eq.~(\ref{eq:condition_twist_custodial}), the $\mathrm{U}(1)_Y$ gauge interaction and the Yukawa interaction violate the custodial symmetry.
It means that the physical consequences of the custodial symmetry violation relate to the mass difference of the W and Z bosons, and the up-type and down-type fermions.
In the following subsections, the physical observables, which are induced by the custodial symmetry violation, are introduced.

Before that, we discuss the conditions for the CP symmetry in the potential.
In order to preserve the CP symmetry in the potential, so-called the \textit{real basis}, in which all parameters in the potential are real, must exist~\cite{Gunion:2005ja,Haber:2010bw}.
The scalar doublet $\Phi_2$ has the degree of the freedom for rephasing, so that there are three rephasing invariants in the potential, $Z_5^* Z_6^2$, $Z_5^* Z_7^2$ and $Z_6^* Z_7$.
As a result, we obtain the conditions for the CP conserving potential in the real $Z_5$ basis~\cite{Botella:1994cs,Davidson:2005cw,Gunion:2005ja,Haber:2010bw}:
\begin{align}
  Z_5 \mathrm{Im}[Z_6^2] = Z_5 \mathrm{Im}[Z_7^2] = \mathrm{Im}[Z_6^* Z_7] = 0.
  \label{eq:condition_CPV} 
\end{align}
Therefore, from eqs.~(\ref{eq:condition_custodial}) and (\ref{eq:condition_CPV}), one can see that the custodial symmetry is violated in the CP violating potential.

In the Yukawa sector, the additional Yukawa matrices $\rho^f$ $(f=u,d,e)$ are able to violate the CP symmetry.
For example, in the case of $\rho^d = \rho^e = 0$, the additional rephasing invariants in the model are $Z_5 (\rho^u)^2$, $Z_6 \rho^u$ and $Z_7 \rho^u$, so that all of them have to be real matrices for the CP symmetry~\cite{Haber:2010bw}.

\subsection{Oblique parameters}

The finite corrections to the self energy of the electroweak gauge bosons are parametrized by the oblique parameters, $S$, $T$, and $U$, which was first introduced by Peskin and Takeuchi~\cite{Peskin:1990zt,Peskin:1991sw}.
Especially, the $T$ parameter represents the difference between the mass corrections of the W and Z bosons, and it is related to the $\rho$ parameter~\cite{Veltman:1977kh,Toussaint:1978zm,Chanowitz:1978uj,Lytel:1980zh,Einhorn:1981cy} as $\alpha_{em} T = \rho - 1$ ($= \Delta \rho$).
The $T$ parameter can be expressed by~\cite{ParticleDataGroup:2022pth}
\begin{align}
  \alpha_{em} T = \frac{\Pi_{WW}^{\mathrm{new}} (0)}{m_W^2} -\frac{\Pi_{ZZ}^{\mathrm{new}}(0)}{m_Z^2},
\end{align}
where $\alpha_{em}$ is the fine-structure constant, and $\Pi_{VV}^{\mathrm{new}}(0)$ ($V=W,Z$) are the corrections from new physics to the two-point self energies of the gauge bosons at the zero momentum.
Those oblique parameters have been precisely tested by the electroweak precision measurements, and the observed value of the $\rho$ parameter is very close to unity~\cite{Baak:2012kk,Baak:2014ora,ParticleDataGroup:2022pth}.

In the 2HDM, the additional scalar bosons radiatively contribute to the self energy of the gauge bosons, and those parameters are modified from the value in the SM.
Especially, the $T$ parameter is an important parameter to know the structure of the Higgs potential because it is sensitive to the violation of the custodial symmetry in the 2HDM~\cite{Lytel:1980zh,Toussaint:1978zm,Pomarol:1993mu,Haber:2010bw}.
When the potential has the custodial symmetry, the corrections to the $T$ parameter from the 2HDM are zero, and there are only corrections from the $\mathrm{U}(1)_Y$ and the Yukawa interactions.
However, within the one-loop level analysis, the inverse is not true, i.e. the small $T$ parameter does not mean that the potential has the custodial symmetry.

For example, when we take a limit $Z_6 = 0$, as one can see from eqs.~(\ref{eq:neut_mass}) and (\ref{eq:relation_neutmass_diagmass}), the squared mass matrix for the neutral scalar bosons is diagonalized with $\mathcal{R}_{ij} = \delta_{ij}$, and the relations $m_{H^\pm}^2 - m_{H_2}^2 = (Z_4 + Z_5) v^2 /2$ and $m_{H^\pm}^2 - m_{H_3}^2 = (Z_4 - Z_5) v^2 /2$ are shown.
Therefore, in the real $Z_5$ basis, the conditions for the custodial symmetric potential are given by 
\begin{align}
  &\mathrm{Custodial~symmetry}: \hspace{1.58cm} m_{H^\pm} = m_{H_3}, ~~~~ Z_7^I = 0 \quad (\chi = 0, ~\pi), \label{eq:condition_custodial_align} \\
  &\mathrm{Twisted~custodial~symmetry}: ~ m_{H^\pm} = m_{H_2}, ~~~~ Z_7^R = 0 ~~~ (\chi = \pi/2, ~3 \pi/2).
  \label{eq:condition_twist_custodial_align}
\end{align}
When we denote the new contribution of the $T$ parameter in the 2HDM as $\Delta T$, at the one-loop level, it is given by~\cite{Pomarol:1993mu,Haber:2010bw}
\begin{align}
  \Delta T = \frac{1}{16 \pi m_W^2 \sin^2 \theta_W} \Big( F(m_{H^\pm}^2, m_{H_2}^2) + F(m_{H^\pm}^2, m_{H_3}^2) - F(m_{H_2}^2, m_{H_3}^2) \Big),
\end{align} 
where the function $F$ is defined by 
\begin{align}
  F(a,b) = \frac{1}{2} (a +b) - \frac{ab}{a-b} \ln \frac{a}{b},
\end{align}
and $\theta_W$ is the Weinberg angle.
Even if $Z_7 \neq 0$, by taking $m_{H_2} = m_{H^\pm}$ or $m_{H_3} = m_{H^\pm}$, $\Delta T$ can vanish at the one-loop level.
This is because, at this order, the $T$ parameter does not depend on $Z_7$, which is not related to the mass formulae of the scalar bosons at the tree level.
In the next subsection, we introduce another independent observable, the $H^\pm W^\mp Z$ vertices, which are radiatively induced by the violation of the custodial symmetry.

As we discussed in the former subsection, the custodial symmetry is violated by the CP violating potential.
Since the $T$ parameter does not depend on $Z_7$ at the one-loop level, the proposition that the potential violates the CP symmetry $\Rightarrow$ $\Delta T \neq 0$, and its inverse are not true.
The CP violating effects to $\Delta T$ from the non-zero $Z_6$ has been discussed in refs.~\cite{Pomarol:1993mu,Haber:2010bw}.

\subsection{Loop induced $H^\pm W^\mp Z$ vertices}

We define the $H^\pm W^\mp Z$ vertices as
\begin{align}
    m_W g V_{\mu \nu}^{\pm} = 
    \begin{minipage}[b]{0.25\linewidth}
    \centering
    \setlength{\feynhandlinesize}{0.7pt}
    \setlength{\feynhandarrowsize}{4pt}
    \begin{tikzpicture}[baseline=-0.15cm]
    \begin{feynhand}
    \vertex (a) at (0,0){$H^\pm$}; \vertex[NWblob] (d) at (1.5,0) {};
    \vertex (e) at (2.2,-1.3){$Z_\nu$};\vertex (f) at (2.2,1.3){$W^\pm_\mu$};
    \propag[chasca, mom={$k_1$}] (a) to (d);
    \propag[bos, mom={$k_3$}] (d) to (e);
    \propag[chabos, mom={$k_2$}] (d) to (f);
    \end{feynhand}
    \end{tikzpicture}
    \end{minipage},
\end{align}
where $k_1$ ($k_2$ and $k_3$) is the ingoing (outgoing) momentum.
Subsequently, $V_{\mu \nu}^{\pm}$ can be decomposed by
\begin{align}
  V_{\mu\nu}^{\pm} = F_{\pm} g_{\mu \nu} + \frac{G_{\pm}}{m_W^2} k^3_\mu k^2_\nu  + \frac{H_{\pm}}{m_W^2} \epsilon_{\mu \nu \rho \sigma}k_3^\rho k_2^\sigma,
  \label{eq:tensor_decomposition}
\end{align}
where $\epsilon_{\mu \nu \rho \sigma}$ is the completely antisymmetric tensor and the external W and Z bosons satisfy the on-shell conditions for $\partial_\mu W^\mu = 0$ and $\partial_\mu Z^\mu = 0$, respectively.
We distinguish $V^+_{\mu \nu}$ and $V^-_{\mu \nu}$ for the discussion of the CP violation.
These vertices come from the following effective operators,
\begin{align}
  \mathrm{Tr}[\sigma_3 (D_\mu \mathbb{M}_1)^\dagger (D^\mu \mathbb{M}_2) ],~~\mathrm{Tr}[\sigma_3 \mathbb{M}_1^\dagger \mathbb{M}_2 F_Z^{\mu\nu} F^W_{\mu \nu} ], ~~ i \epsilon_{\mu \nu \rho \sigma} \mathrm{Tr}[\sigma_3 \mathbb{M}_1^\dagger \mathbb{M}_2 F_Z^{\mu\nu} F^W_{\rho \sigma} ], 
  \label{eq:effective_operator}
\end{align}
and the hermitian conjugate of them, where $F_W^{\mu\nu}$ and $F_Z^{\mu \nu}$ are the field tensors of the W and Z bosons, respectively.
All of these operators violate the custodial $\mathrm{SU}(2)_V$ symmetry.
In the 2HDM, these effective operators do not exist at the tree level, however, when the Lagrangian violates the custodial symmetry, these are induced by the loop effects.
Especially, the first operator in eq.~(\ref{eq:effective_operator}), which corresponds to the $F_{\pm}$ term in eq.~(\ref{eq:tensor_decomposition}), is the dimension-four operator, so that non-decoupling quantum effects of the additional scalar bosons are enhanced~\cite{Kanemura:1997ej}.
On the other hand, the second and third operators, which are dimension-six, are suppressed by the quadratic scale of the heavy particles.
Therefore, the $F_{\pm}$ term contribution is important to discuss the quantum effects of the additional heavy scalar bosons.
We note that the third operator also breaks the parity symmetry which is conserved in the bosonic sector, so that only the fermion-loop contributions cause non-zero $H_{\pm}$.

Although it has also been known that the $H^\pm W^\mp \gamma$ vertices are generated at the loop level in the 2HDM~\cite{Gunion:1987ke,CapdequiPeyranere:1990qk,Raychaudhuri:1992tm,Raychaudhuri:1994dc,Diaz-Cruz:2001thx,Hernandez-Sanchez:2004yid,BarradasGuevara:2010xs}, due to the Ward--Takahashi identity $p^\mu_\gamma V_{\mu \nu} = 0$, the $F_\pm$ term must vanish.
In this sense, a large enhancement by the heavy scalar bosons cannot be expected, compared with the $H^\pm W^\mp Z$ vertices.
In this paper, we focus on the $H^\pm W^\mp Z$ vertices to see the quantum effects of the additional scalar bosons.

When $m_{H^\pm} > m_W +m_Z$, the decays $H^\pm \to W^\pm Z$ through these vertices are kinematically allowed.
The decay rates of $H^\pm \to W^\pm Z$ are given by
\begin{align}
  \Gamma (H^\pm \to W^\pm Z) = \frac{m_{H^\pm} \lambda^{\frac{1}{2}}(1,w,z) }{16 \pi} \Big( |\mathcal{M}_{LL}|^2 +|\mathcal{M}_{TT}|^2  \Big),
\end{align}
where $w = m_W^2 / m_{H^\pm}^2$, $z = m_Z^2 / m_{H^\pm}^2$ and $\lambda (a,b,c) = (a-b-c)^2 - 4 bc$.
The amplitudes $\mathcal{M}_{LL}$ and $\mathcal{M}_{TT}$ are the contributions from the longitudinal and transverse modes of the external gauge bosons, respectively.
These amplitudes are given by 
\begin{align}
  &|\mathcal{M}_{LL}|^2 = \frac{g^2}{4 z} \bigg| (1-w-z)F_{\pm} + \frac{\lambda(1,w,z)}{2w} G_{\pm} \bigg|^2, \notag \\
  &|\mathcal{M}_{TT}|^2 = g^2 \bigg( 2w |F_{\pm}|^2 + \frac{\lambda(1,w,z)}{2w} |H_{\pm}|^2 \bigg).
\end{align}

\begin{figure}[t]
  \begin{minipage}[b]{0.45\linewidth}
  \centering
  \setlength{\feynhandlinesize}{0.7pt}
  \setlength{\feynhandarrowsize}{4pt}
  \setlength{\feynhanddotsize}{2mm}
  \begin{tikzpicture}
  \begin{feynhand}
  \vertex (a) at (0,0){$H^+$}; \vertex (b) at (3,-1){$G^0$};
  \vertex (c) at (3,1){$G^+$}; \vertex [dot] (d) at (1,0) {};
  \vertex (e) at (2,-1);\vertex [ringdot] (f) at (2,1) {};
  \propag[chasca] (a) to (d);
  \propag[sca] (d) to [edge label' = $H_2$] (e);
  \propag[sca] (e) to [edge label' = $H_3$] (f);
  \propag[chasca] (f) to (c);
  \propag[sca] (e) to (b);
  \propag[chasca] (d) to [edge label = $H^\pm$] (f);
  \end{feynhand}
  \end{tikzpicture}
  ~~
  \begin{tikzpicture}
      \begin{feynhand}
      \vertex (a) at (0,0){$H^+$}; \vertex (b) at (3,-0.8){$G^0$};
      \vertex (c) at (3,0.8){$G^+$}; \vertex [dot] (d) at (1,0) {};
      \vertex [ringdot] (e) at (2,0) {};
      \propag[chasca] (a) to (d);
      \propag[chasca] (d) to [half left, looseness=1.6, edge label = $H^\pm$] (e);
      \propag[sca] (d) to [half right, looseness=1.6, edge label' = $H_2$] (e);
      \propag[chasca] (e) to (c);
      \propag[sca] (e) to (b);
      \end{feynhand}
  \end{tikzpicture}
  ~~
  \begin{tikzpicture}
      \begin{feynhand}
      \vertex (a) at (0,0){$H^+$}; \vertex (b) at (1.5,-1){$G^0$};
      \vertex (c) at (3,0.8){$G^+$}; \vertex [dot] (d) at (1,0) {};
      \vertex [ringdot] (e) at (1.8,0.5) {};
      \propag[chasca] (a) to (d);
      \propag[chasca] (d) to [half left, looseness=1.5, edge label = $H^\pm$] (e);
      \propag[sca] (d) to [half right, looseness=1.5, edge label' = $H_3$] (e);
      \propag[chasca] (e) to (c);
      \propag[sca] (d) to (b);
      \end{feynhand}
  \end{tikzpicture}
  \end{minipage}
  \begin{minipage}[b]{0.45\linewidth}
      \centering
      \setlength{\feynhandlinesize}{0.7pt}
      \setlength{\feynhandarrowsize}{4pt}
      \setlength{\feynhanddotsize}{2mm}
      \begin{tikzpicture}
      \begin{feynhand}
      \vertex (a) at (0,0){$H^+$}; \vertex (b) at (3,-1){$G^0$};
      \vertex (c) at (3,1){$G^+$}; \vertex [ringdot] (d) at (1,0) {};
      \vertex (e) at (2,-1);\vertex [dot] (f) at (2,1) {};
      \propag[chasca] (a) to (d);
      \propag[sca] (d) to [edge label' = $H_3$] (e);
      \propag[sca] (e) to [edge label' = $H_2$] (f);
      \propag[chasca] (f) to (c);
      \propag[sca] (e) to (b);
      \propag[chasca] (d) to [edge label = $H^\pm$] (f);
      \end{feynhand}
      \end{tikzpicture}
      ~~
      \begin{tikzpicture}
          \begin{feynhand}
          \vertex (a) at (0,0){$H^+$}; \vertex (b) at (3,-0.8){$G^0$};
          \vertex (c) at (3,0.8){$G^+$}; \vertex [ringdot] (d) at (1,0) {};
          \vertex [dot] (e) at (2,0) {};
          \propag[chasca] (a) to (d);
          \propag[chasca] (d) to [half left, looseness=1.6, edge label = $H^\pm$] (e);
          \propag[sca] (d) to [half right, looseness=1.6, edge label' = $H_3$] (e);
          \propag[chasca] (e) to (c);
          \propag[sca] (e) to (b);
          \end{feynhand}
      \end{tikzpicture}
      ~~
      \begin{tikzpicture}
          \begin{feynhand}
          \vertex (a) at (0,0){$H^+$}; \vertex (b) at (1.5,-1){$G^0$};
          \vertex (c) at (3,0.8){$G^+$}; \vertex [ringdot] (d) at (1,0) {};
          \vertex [dot] (e) at (1.8,0.5) {};
          \propag[chasca] (a) to (d);
          \propag[chasca] (d) to [half left, looseness=1.5, edge label = $H^\pm$] (e);
          \propag[sca] (d) to [half right, looseness=1.5, edge label' = $H_2$] (e);
          \propag[chasca] (e) to (c);
          \propag[sca] (d) to (b);
          \end{feynhand}
      \end{tikzpicture}
      \end{minipage}
      \caption{The scalar contributions of the decay $H^+ \to G^+ G^0$ under the condition $Z_6=0$.
      The white (black) dots express the interaction which violates the (twisted) custodial symmetry.}
      \label{fig:HWZ_ET_ALIGN}
\end{figure}
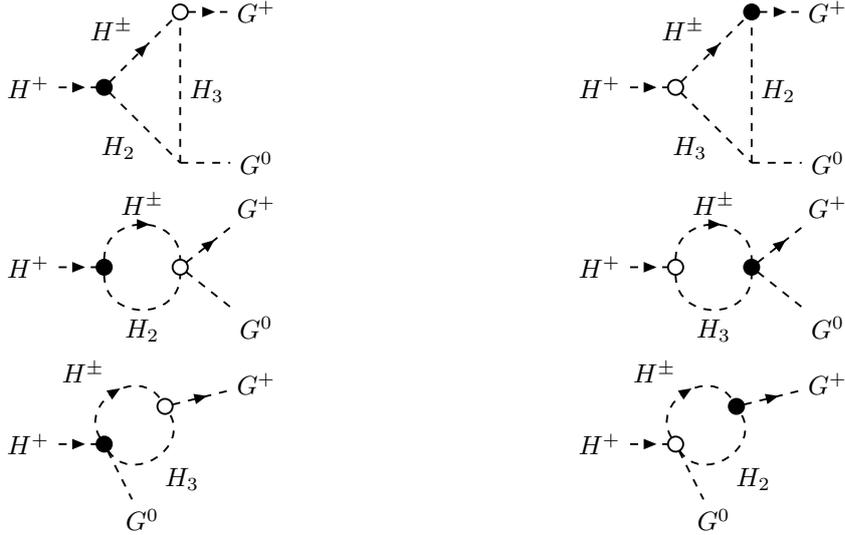

Before closing this section, we discuss the relation among the decays $H^\pm \to W^\pm Z$, the custodial symmetry, and the CP symmetry in our model.
In section~\ref{sec:vertex} and appendix~\ref{sec:app1}, we will give all diagrams and the full formulae for $F_{\pm}$, $G_{\pm}$ and $H_{\pm}$ in the general 2HDM.
However, in order to see the essence, we here take the heavy mass limit, where $m_{W,Z} / m_{H^\pm} \to 0$.
In this limit, the longitudinal contribution $\mathcal{M}_{LL}$ is dominant in the decays $H^\pm \to W^\pm Z$.
In addition, thanks to the equivalence theorem~\cite{Cornwall:1974km,Lee:1977eg}, the calculation of $\mathcal{M}_{LL}$ can be much simpler than the original calculation by replacing the external gauge bosons to the corresponding NG bosons.

As same as the discussion of the $T$ parameter, we here take the $Z_6 = 0$ condition just for simplicity.
\footnote{Again, in section~\ref{sec:vertex} and appendix~\ref{sec:app1}, the full calculation and numerical results including $Z_6 \neq 0$ case without using the equivalence theorem are shown.}
The diagrams which contribute to $\Gamma (H^+ \to G^+ G^0)$ from the potential are shown in figure~\ref{fig:HWZ_ET_ALIGN}.
Here we have employed the Landau gauge and calculated in the real $Z_5$ basis.
The white (black) dots in the figure~\ref{fig:HWZ_ET_ALIGN} are proportional to $Z_7^I$ ($Z_7^R$) or $m_{H^\pm}^2-m_{H_3}^2$ ($m_{H^\pm}^2-m_{H_2}^2$), so that they break the (twisted) custodial symmetry, as it can be understood from eq.~(\ref{eq:condition_custodial_align}).
At the one-loop level, the amplitudes are given by 
\begin{align}
  \mathcal{M}(H^\pm \to G^\pm G^0) &= \frac{\pm i Z_7^R}{16 \pi^2 v} (m_{H^\pm}^2 - m_{H_3}^2) \Big\{ (m_{H_2}^2 - m_{H_3}^2) C_0[-k_2,-k_3;m_{H^\pm}^2,m_{H_3}^2,m_{H_2}^2] \notag \\
  &+ B_0 [-k_2; m_{H^\pm}^2, m_{H_3}^2] -B_0 [-k_1; m_{H^\pm}^2, m_{H_2}^2] \Big\} \notag \\
  &+\frac{Z_7^I}{16 \pi^2 v} (m_{H^\pm}^2 - m_{H_2}^2) \Big\{ (m_{H_2}^2 - m_{H_3}^2) C_0[-k_2,-k_3;m_{H^\pm}^2,m_{H_2}^2,m_{H_3}^2] \notag \\
  &- B_0 [-k_2; m_{H^\pm}^2, m_{H_2}^2] +B_0 [-k_1; m_{H^\pm}^2, m_{H_3}^2] \Big\},
  \label{eq:amplitude_ET}
\end{align}
where the $B_0$ and $C_0$ functions are the scalar Passarino--Veltmann functions~\cite{tHooft:1978jhc,Passarino:1978jh}, whose definitions are summarized in appendix~\ref{sec:app2}.
We can see the equivalence of the custodial and the twisted custodial symmetry from eq.~(\ref{eq:amplitude_ET}).
When the conditions for the (twisted) custodial symmetry, $Z_7^I = 0$ and $m_{H^\pm}^2 = m_{H_3}^2$ ($Z_7^R = 0$ and $m_{H^\pm}^2 = m_{H_2}^2$), are satisfied, the amplitude is $\mathcal{M}(H^\pm \to G^\pm G^0) = 0$.
Due to the CP violating potential, there are two terms which are proportional to the real or imaginary part of $Z_7$.
In the CP conserving 2HDM, there are no corresponding coupling constants to $Z_7^I$ in the Higgs basis.
Therefore, the second term in eq.~(\ref{eq:amplitude_ET}) is the new part from the CP violating potential.
We note that, however, the existence of only the second term does not mean the CP violation in the potential.
This is because, under the $Z_6 =0$ condition, the remaining rephasing invariant is $Z_5^* Z_7^2$, so that $\mathrm{Im}[Z_7^2] \propto Z_7^R Z_7^I$ dependence in the real $Z_5$ basis is needed for measuring the CP violation.
Actually, the loop functions in eq.~(\ref{eq:amplitude_ET}) are always real, so that both squared amplitudes for $H^+$ and $H^-$ take forms as $(Z_7^R)^2 c + (Z_7^I)^2 c^\prime$, where $c$ and $c^\prime$ are real constants.
As a result, the decays $H^\pm \to W^\pm Z$ cannot be the observable to measure $\mathrm{Im}[Z_7^2]$.

Even in the general formulae which are shown in appendix~\ref{sec:app1}, this feature is not changed.
However, under the existence of the fermion-loop contributions with the additional Yukawa matrices $\rho^f$, an asymmetry between the decay rates of $H^+ \to W^+ Z$ and $H^- \to W^- Z$ can appear.
This is because the loop functions in those contributions can have imaginary parts, and $\mathrm{Im}[Z_7 \rho^f]$ related differences appear in these decay rates.
In the following sections, we also give discussions and results for the CP asymmetry in these decays.

\section{The analyses for the decays $H^\pm \to W^\pm Z$ in the general two Higgs doublet model \label{sec:vertex}}

In this section, we discuss the decays $H^\pm \to W^\pm Z$ in the general 2HDM.
We calculate the decay in the t'Hooft--Feynman gauge at the one-loop level.
The full analytic expressions for the $H^\pm W^\mp Z$ vertices are given in appendix~\ref{sec:app1}.
All numerical results which are shown in the following subsections are given by using these formulae.

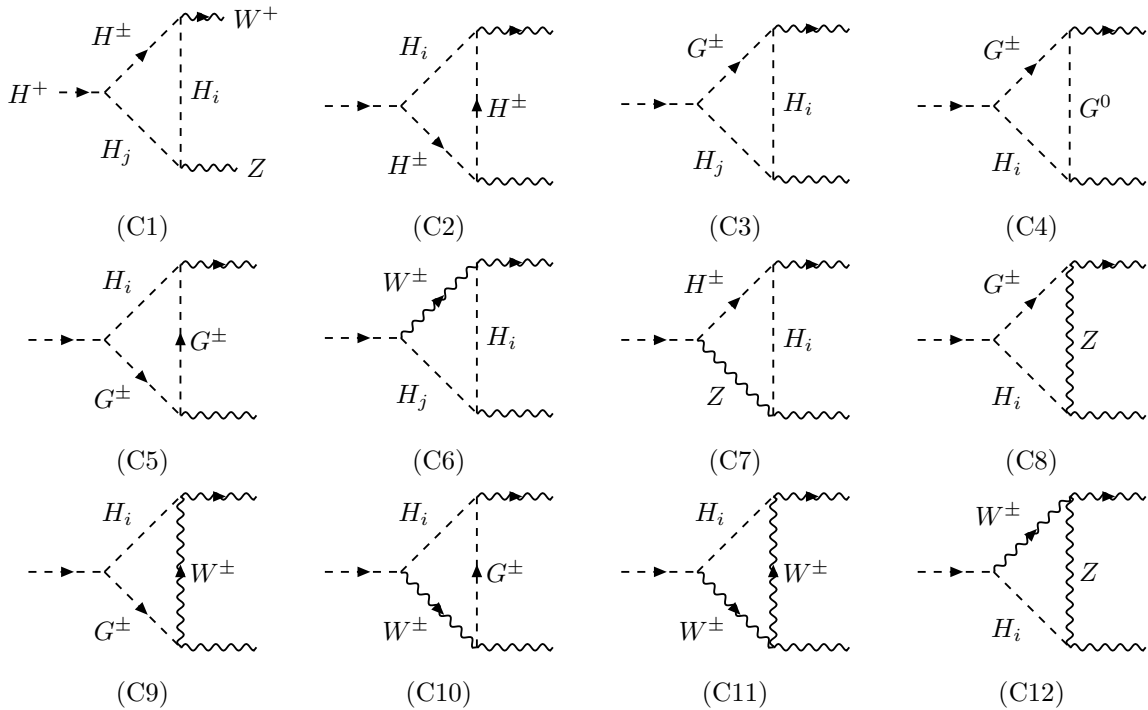
\begin{figure}[t]
  \centering
  \begin{minipage}[b]{0.23\linewidth}
  \centering
  \setlength{\feynhandlinesize}{0.7pt}
  \setlength{\feynhandarrowsize}{4pt}
  \begin{tikzpicture}
  \begin{feynhand}
  \vertex (a) at (0,0){$H^+$}; \vertex (b) at (3,-1){$Z$};
  \vertex (c) at (3,1){$W^+$}; \vertex (d) at (1,0);
  \vertex (e) at (2,-1);\vertex (f) at (2,1);
  \propag[chasca] (a) to (d);
  \propag[sca] (d) to [edge label' = $H_j$] (e);
  \propag[sca] (e) to [edge label' = $H_i$] (f);
  \propag[chabos] (f) to (c);
  \propag[bos] (e) to (b);
  \propag[chasca] (d) to [edge label = $H^\pm$] (f);
  \end{feynhand}
  \end{tikzpicture}
  \subcaption*{(C1)}
  \end{minipage}
  \begin{minipage}[b]{0.23\linewidth}
      \centering
      \setlength{\feynhandlinesize}{0.7pt}
      \setlength{\feynhandarrowsize}{4pt}
      \begin{tikzpicture}
      \begin{feynhand}
      \vertex (a) at (0,0); \vertex (b) at (3,-1);
      \vertex (c) at (3,1); \vertex (d) at (1,0);
      \vertex (e) at (2,-1);\vertex (f) at (2,1);
      \propag[chasca] (a) to (d);
      \propag[chasca] (d) to [edge label' = $H^\pm$] (e);
      \propag[chasca] (e) to [edge label' = $H^\pm$] (f);
      \propag[chabos] (f) to (c);
      \propag[bos] (e) to (b);
      \propag[sca] (d) to [edge label = $H_i$] (f);
      \end{feynhand}
      \end{tikzpicture}
      \subcaption*{(C2)}
      \end{minipage}
  \begin{minipage}[b]{0.23\linewidth}
      \centering
      \setlength{\feynhandlinesize}{0.7pt}
      \setlength{\feynhandarrowsize}{4pt}
      \begin{tikzpicture}
      \begin{feynhand}
      \vertex (a) at (0,0); \vertex (b) at (3,-1);
      \vertex (c) at (3,1); \vertex (d) at (1,0);
      \vertex (e) at (2,-1);\vertex (f) at (2,1);
      \propag[chasca] (a) to (d);
      \propag[sca] (d) to [edge label' = $H_j$] (e);
      \propag[sca] (e) to [edge label' = $H_i$] (f);
      \propag[chabos] (f) to (c);
      \propag[bos] (e) to (b);
      \propag[chasca] (d) to [edge label = $G^\pm$] (f);
      \end{feynhand}
      \end{tikzpicture}
      \subcaption*{(C3)}
      \end{minipage}
  \begin{minipage}[b]{0.23\linewidth}
      \centering
      \setlength{\feynhandlinesize}{0.7pt}
      \setlength{\feynhandarrowsize}{4pt}
      \begin{tikzpicture}
      \begin{feynhand}
      \vertex (a) at (0,0); \vertex (b) at (3,-1);
      \vertex (c) at (3,1); \vertex (d) at (1,0);
      \vertex (e) at (2,-1);\vertex (f) at (2,1);
      \propag[chasca] (a) to (d);
      \propag[sca] (d) to [edge label' = $H_i$] (e);
      \propag[sca] (e) to [edge label' = $G^0$] (f);
      \propag[chabos] (f) to (c);
      \propag[bos] (e) to (b);
      \propag[chasca] (d) to [edge label = $G^\pm$] (f);
      \end{feynhand}
      \end{tikzpicture}
      \subcaption*{(C4)}
      \end{minipage}
  \begin{minipage}[b]{0.23\linewidth}
      \centering
      \setlength{\feynhandlinesize}{0.7pt}
      \setlength{\feynhandarrowsize}{4pt}
      \begin{tikzpicture}
      \begin{feynhand}
      \vertex (a) at (0,0); \vertex (b) at (3,-1);
      \vertex (c) at (3,1); \vertex (d) at (1,0);
      \vertex (e) at (2,-1);\vertex (f) at (2,1);
      \propag[chasca] (a) to (d);
      \propag[chasca] (d) to [edge label' = $G^\pm$] (e);
      \propag[chasca] (e) to [edge label' = $G^\pm$] (f);
      \propag[chabos] (f) to (c);
      \propag[bos] (e) to (b);
      \propag[sca] (d) to [edge label = $H_i$] (f);
      \end{feynhand}
      \end{tikzpicture}
      \subcaption*{(C5)}
      \end{minipage}
  \begin{minipage}[b]{0.23\linewidth}
      \centering
      \setlength{\feynhandlinesize}{0.7pt}
      \setlength{\feynhandarrowsize}{4pt}
      \begin{tikzpicture}
      \begin{feynhand}
      \vertex (a) at (0,0); \vertex (b) at (3,-1);
      \vertex (c) at (3,1); \vertex (d) at (1,0);
      \vertex (e) at (2,-1);\vertex (f) at (2,1);
      \propag[chasca] (a) to (d);
      \propag[sca] (d) to [edge label' = $H_j$] (e);
      \propag[sca] (e) to [edge label' = $H_i$] (f);
      \propag[chabos] (f) to (c);
      \propag[bos] (e) to (b);
      \propag[chabos] (d) to [edge label = $W^\pm$] (f);
      \end{feynhand}
      \end{tikzpicture}
      \subcaption*{(C6)}
      \end{minipage}
  \begin{minipage}[b]{0.23\linewidth}
      \centering
      \setlength{\feynhandlinesize}{0.7pt}
      \setlength{\feynhandarrowsize}{4pt}
      \begin{tikzpicture}
      \begin{feynhand}
      \vertex (a) at (0,0); \vertex (b) at (3,-1);
      \vertex (c) at (3,1); \vertex (d) at (1,0);
      \vertex (e) at (2,-1);\vertex (f) at (2,1);
      \propag[chasca] (a) to (d);
      \propag[bos] (d) to [edge label' = $Z$] (e);
      \propag[sca] (e) to [edge label' = $H_i$] (f);
      \propag[chabos] (f) to (c);
      \propag[bos] (e) to (b);
      \propag[chasca] (d) to [edge label = $H^\pm$] (f);
      \end{feynhand}
      \end{tikzpicture}
      \subcaption*{(C7)}
      \end{minipage}
  \begin{minipage}[b]{0.23\linewidth}
      \centering
      \setlength{\feynhandlinesize}{0.7pt}
      \setlength{\feynhandarrowsize}{4pt}
      \begin{tikzpicture}
      \begin{feynhand}
      \vertex (a) at (0,0); \vertex (b) at (3,-1);
      \vertex (c) at (3,1); \vertex (d) at (1,0);
      \vertex (e) at (2,-1);\vertex (f) at (2,1);
      \propag[chasca] (a) to (d);
      \propag[sca] (d) to [edge label' = $H_i$] (e);
      \propag[bos] (e) to [edge label' = $Z$] (f);
      \propag[chabos] (f) to (c);
      \propag[bos] (e) to (b);
      \propag[chasca] (d) to [edge label = $G^\pm$] (f);
      \end{feynhand}
      \end{tikzpicture}
      \subcaption*{(C8)}
      \end{minipage}
  \begin{minipage}[b]{0.23\linewidth}
      \centering
      \setlength{\feynhandlinesize}{0.7pt}
      \setlength{\feynhandarrowsize}{4pt}
      \begin{tikzpicture}
      \begin{feynhand}
      \vertex (a) at (0,0); \vertex (b) at (3,-1);
      \vertex (c) at (3,1); \vertex (d) at (1,0);
      \vertex (e) at (2,-1);\vertex (f) at (2,1);
      \propag[chasca] (a) to (d);
      \propag[chasca] (d) to [edge label' = $G^\pm$] (e);
      \propag[chabos] (e) to [edge label' = $W^\pm$] (f);
      \propag[chabos] (f) to (c);
      \propag[bos] (e) to (b);
      \propag[sca] (d) to [edge label = $H_i$] (f);
      \end{feynhand}
      \end{tikzpicture}
      \subcaption*{(C9)}
      \end{minipage}
  \begin{minipage}[b]{0.23\linewidth}
      \centering
      \setlength{\feynhandlinesize}{0.7pt}
      \setlength{\feynhandarrowsize}{4pt}
      \begin{tikzpicture}
      \begin{feynhand}
      \vertex (a) at (0,0); \vertex (b) at (3,-1);
      \vertex (c) at (3,1); \vertex (d) at (1,0);
      \vertex (e) at (2,-1);\vertex (f) at (2,1);
      \propag[chasca] (a) to (d);
      \propag[chabos] (d) to [edge label' = $W^\pm$] (e);
      \propag[chasca] (e) to [edge label' = $G^\pm$] (f);
      \propag[chabos] (f) to (c);
      \propag[bos] (e) to (b);
      \propag[sca] (d) to [edge label = $H_i$] (f);
      \end{feynhand}
      \end{tikzpicture}
      \subcaption*{(C10)}
  \end{minipage}
  \begin{minipage}[b]{0.23\linewidth}
      \centering
      \setlength{\feynhandlinesize}{0.7pt}
      \setlength{\feynhandarrowsize}{4pt}
      \begin{tikzpicture}
      \begin{feynhand}
      \vertex (a) at (0,0); \vertex (b) at (3,-1);
      \vertex (c) at (3,1); \vertex (d) at (1,0);
      \vertex (e) at (2,-1);\vertex (f) at (2,1);
      \propag[chasca] (a) to (d);
      \propag[chabos] (d) to [edge label' = $W^\pm$] (e);
      \propag[chabos] (e) to [edge label' = $W^\pm$] (f);
      \propag[chabos] (f) to (c);
      \propag[bos] (e) to (b);
      \propag[sca] (d) to [edge label = $H_i$] (f);
      \end{feynhand}
      \end{tikzpicture}
      \subcaption*{(C11)}
  \end{minipage}
  \begin{minipage}[b]{0.23\linewidth}
      \centering
      \setlength{\feynhandlinesize}{0.7pt}
      \setlength{\feynhandarrowsize}{4pt}
      \begin{tikzpicture}
      \begin{feynhand}
      \vertex (a) at (0,0); \vertex (b) at (3,-1);
      \vertex (c) at (3,1); \vertex (d) at (1,0);
      \vertex (e) at (2,-1);\vertex (f) at (2,1);
      \propag[chasca] (a) to (d);
      \propag[sca] (d) to [edge label' = $H_i$] (e);
      \propag[bos] (e) to [edge label' = $Z$] (f);
      \propag[chabos] (f) to (c);
      \propag[bos] (e) to (b);
      \propag[chabos] (d) to [edge label = $W^\pm$] (f);
      \end{feynhand}
      \end{tikzpicture}
      \subcaption*{(C12)}
  \end{minipage}
  \caption{The $C$ type contributions.
  The indices $i$ and $j$ run $1,2,3$, which denote the mass eigenstates of the neutral scalar bosons.}
  \label{fig:HWZ_full_Ctype}
\end{figure}
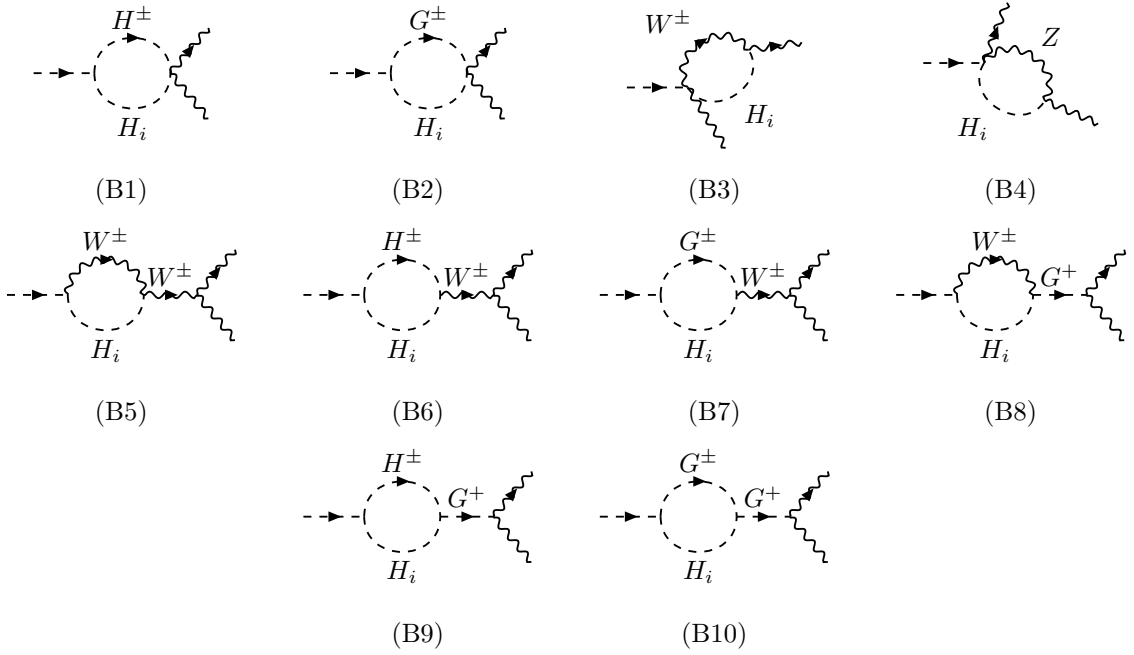
\begin{figure}[h]
  \centering
  \begin{minipage}[b]{0.23\linewidth}
  \setlength{\feynhandlinesize}{0.7pt}
  \setlength{\feynhandarrowsize}{4pt}
  \centering
  \begin{tikzpicture}
      \begin{feynhand}
      \vertex (a) at (0.2,0); \vertex (b) at (2.5,-0.6);
      \vertex (c) at (2.5,0.6); \vertex (d) at (1,0);
      \vertex (e) at (2,0);
      \propag[chasca] (a) to (d);
      \propag[chasca] (d) to [half left, looseness=1.6, edge label = $H^\pm$] (e);
      \propag[sca] (d) to [half right, looseness=1.6, edge label' = $H_i$] (e);
      \propag[chabos] (e) to (c);
      \propag[bos] (e) to (b);
      \end{feynhand}
  \end{tikzpicture}
  \subcaption*{(B1)}
  \end{minipage}
  \begin{minipage}[b]{0.23\linewidth}
      \setlength{\feynhandlinesize}{0.7pt}
      \setlength{\feynhandarrowsize}{4pt}
      \centering
      \begin{tikzpicture}
          \begin{feynhand}
          \vertex (a) at (0.2,0); \vertex (b) at (2.5,-0.6);
          \vertex (c) at (2.5,0.6); \vertex (d) at (1,0);
          \vertex (e) at (2,0);
          \propag[chasca] (a) to (d);
          \propag[chasca] (d) to [half left, looseness=1.6, edge label = $G^\pm$] (e);
          \propag[sca] (d) to [half right, looseness=1.6, edge label' = $H_i$] (e);
          \propag[chabos] (e) to (c);
          \propag[bos] (e) to (b);
          \end{feynhand}
      \end{tikzpicture}
      \subcaption*{(B2)}
  \end{minipage}
  \begin{minipage}[b]{0.23\linewidth}
  \setlength{\feynhandlinesize}{0.7pt}
  \setlength{\feynhandarrowsize}{4pt}
  \centering
  \begin{tikzpicture}
      \begin{feynhand}
      \vertex (a) at (0.2,0); \vertex (b) at (1.5,-0.8);
      \vertex (c) at (2.5,0.6); \vertex (d) at (1,0);
      \vertex (e) at (1.8,0.5);
      \propag[chasca] (a) to (d);
      \propag[chabos] (d) to [half left, looseness=1.5, edge label = $W^\pm$] (e);
      \propag[sca] (d) to [half right, looseness=1.5, edge label' = $H_i$] (e);
      \propag[chabos] (e) to (c);
      \propag[bos] (d) to (b);
      \end{feynhand}
  \end{tikzpicture}
  \subcaption*{(B3)}
  \end{minipage}
  \begin{minipage}[b]{0.23\linewidth}
      \setlength{\feynhandlinesize}{0.7pt}
      \setlength{\feynhandarrowsize}{4pt}
      \centering
      \begin{tikzpicture}
          \begin{feynhand}
          \vertex (a) at (0.2,0); \vertex (b) at (1.3,0.8);
          \vertex (c) at (2.5,-0.8); \vertex (d) at (1,0);
          \vertex (e) at (1.8,-0.5);
          \propag[chasca] (a) to (d);
          \propag[bos] (d) to [half left, looseness=1.5, edge label = $Z$] (e);
          \propag[sca] (d) to [half right, looseness=1.5, edge label' = $H_i$] (e);
          \propag[bos] (e) to (c);
          \propag[chabos] (d) to (b);
          \end{feynhand}
      \end{tikzpicture}
      \subcaption*{(B4)}
  \end{minipage}
  \begin{minipage}[b]{0.23\linewidth}
      \setlength{\feynhandlinesize}{0.7pt}
      \setlength{\feynhandarrowsize}{4pt}
      \centering
      \begin{tikzpicture}
          \begin{feynhand}
          \vertex (a) at (0.2,0); \vertex (b) at (3.2,-0.6);
          \vertex (c) at (3.2,0.6); \vertex (d) at (1,0);
          \vertex (e) at (2,0); \vertex (f) at (2.7,0);
          \propag[chasca] (a) to (d);
          \propag[chabos] (d) to [half left, looseness=1.6, edge label = $W^\pm$] (e);
          \propag[sca] (d) to [half right, looseness=1.6, edge label' = $H_i$] (e);
          \propag[chabos] (f) to (c);
          \propag[bos] (f) to (b);
          \propag[chabos] (e) to [edge label = $W^\pm$] (f);
          \end{feynhand}
      \end{tikzpicture}
      \subcaption*{(B5)}
  \end{minipage}
  \begin{minipage}[b]{0.23\linewidth}
      \setlength{\feynhandlinesize}{0.7pt}
      \setlength{\feynhandarrowsize}{4pt}
      \centering
      \begin{tikzpicture}
          \begin{feynhand}
          \vertex (a) at (0.2,0); \vertex (b) at (3.2,-0.6);
          \vertex (c) at (3.2,0.6); \vertex (d) at (1,0);
          \vertex (e) at (2,0); \vertex (f) at (2.7,0);
          \propag[chasca] (a) to (d);
          \propag[chasca] (d) to [half left, looseness=1.6, edge label = $H^\pm$] (e);
          \propag[sca] (d) to [half right, looseness=1.6, edge label' = $H_i$] (e);
          \propag[chabos] (f) to (c);
          \propag[bos] (f) to (b);
          \propag[chabos] (e) to [edge label = $W^\pm$] (f);
          \end{feynhand}
      \end{tikzpicture}
      \subcaption*{(B6)}
  \end{minipage}
  \begin{minipage}[b]{0.23\linewidth}
      \setlength{\feynhandlinesize}{0.7pt}
      \setlength{\feynhandarrowsize}{4pt}
      \centering
      \begin{tikzpicture}
          \begin{feynhand}
          \vertex (a) at (0.2,0); \vertex (b) at (3.2,-0.6);
          \vertex (c) at (3.2,0.6); \vertex (d) at (1,0);
          \vertex (e) at (2,0); \vertex (f) at (2.7,0);
          \propag[chasca] (a) to (d);
          \propag[chasca] (d) to [half left, looseness=1.6, edge label = $G^\pm$] (e);
          \propag[sca] (d) to [half right, looseness=1.6, edge label' = $H_i$] (e);
          \propag[chabos] (f) to (c);
          \propag[bos] (f) to (b);
          \propag[chabos] (e) to [edge label = $W^\pm$] (f);
          \end{feynhand}
      \end{tikzpicture}
      \subcaption*{(B7)}
  \end{minipage}
  \begin{minipage}[b]{0.23\linewidth}
      \setlength{\feynhandlinesize}{0.7pt}
      \setlength{\feynhandarrowsize}{4pt}
      \centering
      \begin{tikzpicture}
          \begin{feynhand}
          \vertex (a) at (0.2,0); \vertex (b) at (3.2,-0.6);
          \vertex (c) at (3.2,0.6); \vertex (d) at (1,0);
          \vertex (e) at (2,0); \vertex (f) at (2.7,0);
          \propag[chasca] (a) to (d);
          \propag[chabos] (d) to [half left, looseness=1.6, edge label = $W^\pm$] (e);
          \propag[sca] (d) to [half right, looseness=1.6, edge label' = $H_i$] (e);
          \propag[chabos] (f) to (c);
          \propag[bos] (f) to (b);
          \propag[chasca] (e) to [edge label = $G^+$] (f);
          \end{feynhand}
      \end{tikzpicture}
      \subcaption*{(B8)}
  \end{minipage}
  \begin{minipage}[b]{0.23\linewidth}
      \setlength{\feynhandlinesize}{0.7pt}
      \setlength{\feynhandarrowsize}{4pt}
      \centering
      \begin{tikzpicture}
          \begin{feynhand}
          \vertex (a) at (0.2,0); \vertex (b) at (3.2,-0.6);
          \vertex (c) at (3.2,0.6); \vertex (d) at (1,0);
          \vertex (e) at (2,0); \vertex (f) at (2.7,0);
          \propag[chasca] (a) to (d);
          \propag[chasca] (d) to [half left, looseness=1.6, edge label = $H^\pm$] (e);
          \propag[sca] (d) to [half right, looseness=1.6, edge label' = $H_i$] (e);
          \propag[chabos] (f) to (c);
          \propag[bos] (f) to (b);
          \propag[chasca] (e) to [edge label = $G^+$] (f);
          \end{feynhand}
      \end{tikzpicture}
      \subcaption*{(B9)}
  \end{minipage}
  \begin{minipage}[b]{0.23\linewidth}
      \setlength{\feynhandlinesize}{0.7pt}
      \setlength{\feynhandarrowsize}{4pt}
      \centering
      \begin{tikzpicture}
          \begin{feynhand}
          \vertex (a) at (0.2,0); \vertex (b) at (3.2,-0.6);
          \vertex (c) at (3.2,0.6); \vertex (d) at (1,0);
          \vertex (e) at (2,0); \vertex (f) at (2.7,0);
          \propag[chasca] (a) to (d);
          \propag[chasca] (d) to [half left, looseness=1.6, edge label = $G^\pm$] (e);
          \propag[sca] (d) to [half right, looseness=1.6, edge label' = $H_i$] (e);
          \propag[chabos] (f) to (c);
          \propag[bos] (f) to (b);
          \propag[chasca] (e) to [edge label = $G^+$] (f);
          \end{feynhand}
      \end{tikzpicture}
      \subcaption*{(B10)}
  \end{minipage}
  \caption{The $B$ type contributions.}
  \label{fig:HWZ_B}
\end{figure}
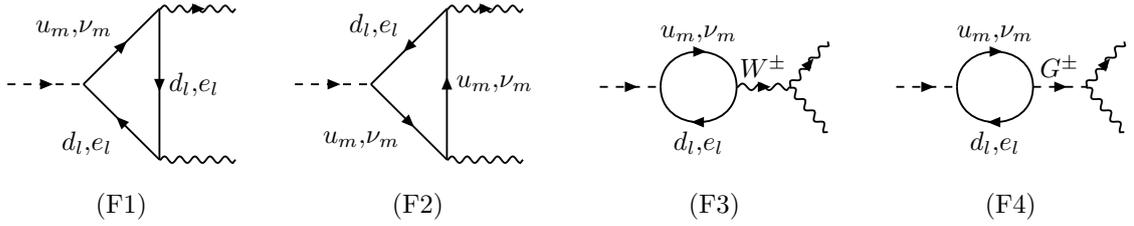
\begin{figure}[h]
  \centering
  \begin{minipage}[b]{0.23\linewidth}
      \centering
      \setlength{\feynhandlinesize}{0.7pt}
      \setlength{\feynhandarrowsize}{4pt}
      \begin{tikzpicture}
          \begin{feynhand}
          \vertex (a) at (0,0); \vertex (b) at (3,-1);
          \vertex (c) at (3,1); \vertex (d) at (1,0);
          \vertex (e) at (2,-1);\vertex (f) at (2,1);
          \propag[chasca] (a) to (d);
          \propag[fer] (e) to [edge label = $d_l\mathrm{,} e_l$] (d);
          \propag[fer] (f) to [edge label = $d_l\mathrm{,} e_l$] (e);
          \propag[chabos] (f) to (c);
          \propag[bos] (e) to (b);
          \propag[fer] (d) to [edge label = $u_m\mathrm{,} \nu_m$] (f);
          \end{feynhand}
          \end{tikzpicture}
      \subcaption*{(F1)}
      \end{minipage}
      \begin{minipage}[b]{0.23\linewidth}
          \centering
          \setlength{\feynhandlinesize}{0.7pt}
          \setlength{\feynhandarrowsize}{4pt}
          \begin{tikzpicture}
          \begin{feynhand}
          \vertex (a) at (0,0); \vertex (b) at (3,-1);
          \vertex (c) at (3,1); \vertex (d) at (1,0);
          \vertex (e) at (2,-1);\vertex (f) at (2,1);
          \propag[chasca] (a) to (d);
          \propag[fer] (d) to [edge label' = $u_m\mathrm{,}\nu_m$] (e);
          \propag[fer] (e) to [edge label' = $u_m\mathrm{,}\nu_m$] (f);
          \propag[chabos] (f) to (c);
          \propag[bos] (e) to (b);
          \propag[fer] (f) to [edge label' = $d_l\mathrm{,}e_l$] (d);
          \end{feynhand}
          \end{tikzpicture}
          \subcaption*{(F2)}
          \end{minipage}
      \begin{minipage}[b]{0.23\linewidth}
          \setlength{\feynhandlinesize}{0.7pt}
          \setlength{\feynhandarrowsize}{4pt}
          \centering
          \begin{tikzpicture}
              \begin{feynhand}
              \vertex (a) at (0.2,0); \vertex (b) at (3.2,-0.6);
              \vertex (c) at (3.2,0.6); \vertex (d) at (1,0);
              \vertex (e) at (2,0); \vertex (f) at (2.7,0);
              \propag[chasca] (a) to (d);
              \propag[fer] (d) to [half left, looseness=1.6, edge label = $u_m\mathrm{,} \nu_m$] (e);
              \propag[fer] (e) to [half left, looseness=1.6, edge label = $d_l\mathrm{,} e_l$] (d);
              \propag[chabos] (f) to (c);
              \propag[bos] (f) to (b);
              \propag[chabos] (e) to [edge label = $W^\pm$] (f);
              \end{feynhand}
          \end{tikzpicture}
          \subcaption*{(F3)}
      \end{minipage}
      \begin{minipage}[b]{0.23\linewidth}
          \setlength{\feynhandlinesize}{0.7pt}
          \setlength{\feynhandarrowsize}{4pt}
          \centering
          \begin{tikzpicture}
              \begin{feynhand}
              \vertex (a) at (0.2,0); \vertex (b) at (3.2,-0.6);
              \vertex (c) at (3.2,0.6); \vertex (d) at (1,0);
              \vertex (e) at (2,0); \vertex (f) at (2.7,0);
              \propag[chasca] (a) to (d);
              \propag[fer] (d) to [half left, looseness=1.6, edge label = $u_m\mathrm{,} \nu_m$] (e);
              \propag[fer] (e) to [half left, looseness=1.6, edge label = $d_l\mathrm{,} e_l$] (d);
              \propag[chabos] (f) to (c);
              \propag[bos] (f) to (b);
              \propag[chasca] (e) to [edge label = $G^\pm$] (f);
              \end{feynhand}
          \end{tikzpicture}
          \subcaption*{(F4)}
      \end{minipage}
  \caption{The fermion contributions.
  The indices $l,m$ denote the fermion flavors.}
  \label{fig:HWZ_full_Fermion}
\end{figure}
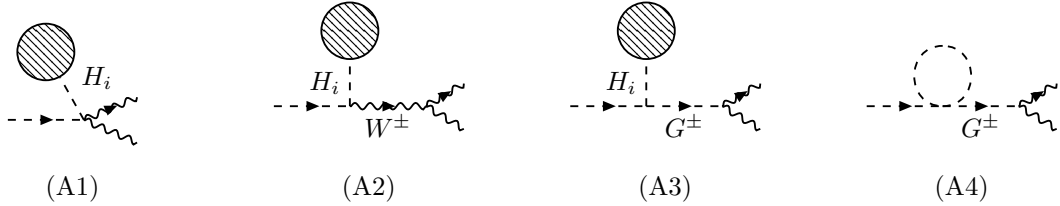
\begin{figure}[h]
  \centering
  \begin{minipage}[b]{0.23\linewidth}
  \setlength{\feynhandlinesize}{0.7pt}
  \setlength{\feynhandarrowsize}{4pt}
  \centering
  \begin{tikzpicture}[baseline=-0.15cm]
      \begin{feynhand}
      \vertex (a) at (0,0); \vertex (d) at (1,0); \vertex[NWblob] (g) at (0.5,0.9) {};
      \vertex (e) at (1.7,-0.3);\vertex (f) at (1.7,0.3);
      \propag[chasca] (a) to (d);
      \propag[bos] (d) to (e);
      \propag[chabos] (d) to (f); \propag[sca] (d) to [edge label' =$H_i$] (g);
      \end{feynhand}
      \end{tikzpicture}
  \subcaption*{(A1)}
  \end{minipage}
  \begin{minipage}[b]{0.23\linewidth}
      \setlength{\feynhandlinesize}{0.7pt}
      \setlength{\feynhandarrowsize}{4pt}
      \centering
      \begin{tikzpicture}[baseline=-0.15cm]
          \begin{feynhand}
          \vertex (a) at (0,0); \vertex (d) at (1,0); \vertex[NWblob] (g) at (1,1) {}; \vertex (h) at (2,0);
          \vertex (e) at (2.5,-0.3);\vertex (f) at (2.5,0.3);
          \propag[chasca] (a) to (d);
          \propag[bos] (h) to (e);
          \propag[chabos] (h) to (f); \propag[sca] (d) to [edge label =$H_i$] (g);
          \propag[chabos] (d) to [edge label' =$W^\pm$] (h);
          \end{feynhand}
          \end{tikzpicture}
      \subcaption*{(A2)}
      \end{minipage}
  \begin{minipage}[b]{0.23\linewidth}
      \setlength{\feynhandlinesize}{0.7pt}
      \setlength{\feynhandarrowsize}{4pt}
      \centering
      \begin{tikzpicture}[baseline=-0.15cm]
          \begin{feynhand}
          \vertex (a) at (0,0); \vertex (d) at (1,0); \vertex[NWblob] (g) at (1,1) {}; \vertex (h) at (2,0);
          \vertex (e) at (2.5,-0.3);\vertex (f) at (2.5,0.3);
          \propag[chasca] (a) to (d);
          \propag[bos] (h) to (e);
          \propag[chabos] (h) to (f); \propag[sca] (d) to [edge label =$H_i$] (g);
          \propag[chasca] (d) to [edge label' =$G^\pm$] (h);
          \end{feynhand}
          \end{tikzpicture}
      \subcaption*{(A3)}
      \end{minipage}
  \begin{minipage}[b]{0.23\linewidth}
      \setlength{\feynhandlinesize}{0.7pt}
      \setlength{\feynhandarrowsize}{4pt}
      \centering
      \begin{tikzpicture}[baseline=-0.15cm]
          \begin{feynhand}
          \vertex (a) at (0,0); \vertex (d) at (1,0); \vertex (g) at (1,0.8); \vertex (h) at (2,0);
          \vertex (e) at (2.5,-0.3);\vertex (f) at (2.5,0.3);
          \propag[chasca] (a) to (d);
          \propag[bos] (h) to (e);
          \propag[chabos] (h) to (f);
          \propag[chasca] (d) to [edge label' =$G^\pm$] (h);
          \propag[sca] (d) to [half right, looseness=1.6] (g);
          \propag[sca] (g) to [half right, looseness=1.6] (d);
          \end{feynhand}
          \end{tikzpicture}
      \subcaption*{(A4)}
      \end{minipage}
  \caption{The $A$ type contributions.}
  \label{fig:HWZ_A}
\end{figure}

In figures~\ref{fig:HWZ_full_Ctype}-\ref{fig:HWZ_A}, all contributions to the decay $H^+ \to W^+ Z$ are shown.
\footnote{The diagrams of the decay $H^- \to W^- Z$ can be obtained by changing the directions of the arrows for the charged particles.}
The boson contributions are shown in figure~\ref{fig:HWZ_full_Ctype} and figure~\ref{fig:HWZ_B}, where $i,j$ $(= 1,2,3)$ are the indices of the neutral scalar bosons in the mass eigenstate.
The diagrams given in figure~\ref{fig:HWZ_full_Ctype} and figure~\ref{fig:HWZ_B} can be expressed by the $C$ and $B$ type Passarino--Veltmann functions, respectively.
The fermion contributions are shown in figure~\ref{fig:HWZ_full_Fermion}, where $l,m$ $(= 1,2,3)$ are the fermion-flavor indices.
Only the fermion contributions give the $H_{\pm}$ term in eq.~(\ref{eq:tensor_decomposition}) because the SM fermions have the chiral interaction.
There are other contributions from the diagrams of the tadpole and the $H^+ G^-$ mixing, which are shown in figure~\ref{fig:HWZ_A}.
These diagrams are expressed by the $A_0$ scalar one-point function. 
Analytic expressions for all diagrams are given in appendix~\ref{sec:app1}.

Here we give a comment on a method for renormalization.
In our model, the effective operators which cause the $H^\pm W^\mp Z$ vertices are absent at the tree level, so that such a one-loop-induced vertex must be finite in the renormalizable theory.
By summing all of those contributions, the divergences in each diagram, which are also shown in appendix~\ref{sec:app1}, cancel out.
Therefore, a renormalization method to absorb these divergences in these vertices is not needed.
However, once a renormalization scheme is adopted and the counter terms in this scheme are fixed, they might give finite corrections to these vertices at the one-loop level.
We expect that those corrections are small, and we do not specify the renormalization scheme in this paper.

\subsection{Model parameters and constraints}
First, for numerical analysis, we explain a method to determine the input parameters.
The free independent parameters in our model are $Y_2^2, Z_{1,2,3,4,5}, Z_6^R, Z_6^I, Z_7^R, Z_7^I$, and $\rho^f$ ($f=u,d,e$).
We take the mixing angles $\alpha_i$ and the mass $m_{H_i}$ ($i=1,2,3$) as the input parameters.
Therefore, the related coupling constants $Z_1, Z_4, Z_5, Z_6^R$ and $Z_6^I$ are determined by
\begin{align}
  \mathcal{M}^2 = \mathcal{R}^T \mathrm{diag}(m_{H_1}^2,m_{H_2}^2,m_{H_3}^2 ) \mathcal{R}.
  \label{eq:relationRandM}
\end{align}
where $\mathcal{M}^2$ is given in eq.~(\ref{eq:neut_mass}).
In order to take the real $Z_5$ basis, a set of input parameters $(\alpha_1, \alpha_2, \alpha_3, m_{H_1}, m_{H_2}, m_{H_3})$ must satisfy $(\mathcal{M}^2)_{23} = (\mathcal{M}^2)_{32} = 0$.
First we determine $(\alpha_1, \alpha_2, m_{H_1}, m_{H_2}, m_{H_3})$ and then numerically get $\alpha_3 = \alpha_3^*$, which satisfies $(\mathcal{M}^2)_{23} = (\mathcal{M}^2)_{32} = 0$.
In the following numerical studies, we choose $\alpha_3^*$ such that the coupling $Z_6$ satisfies $0 \le \mathrm{arg}[Z_6] < \pi/2$.
We also treat $m_{H^\pm}$ and $Y_2^2$ as the input parameters and fix $-Y_2^2 = (20~\mathrm{GeV})^2$.
The coupling constant $Z_2$, which is irrelevant to the decays $H^\pm \to W^\pm Z$ at the one-loop level, is set by $Z_2 = 2$.

Second, we have the general complex Yukawa matrices $\rho^f$ in our model. 
Although the formulae of the $H^\pm W^\mp Z$ vertices with general $\rho^f$ are shown in appendix~\ref{sec:app1}, the contributions from each flavor give similar results.
Under the data from the flavor experiments, $\rho_{tt}$ and $\rho_{tc}$, which are $(3,3)$ and $(3,2)$ elements in $\rho^u$, respectively, are normally allowed to be larger than the other additional Yukawa coupling constants~\cite{Crivellin:2013wna}.
In addition, effects from $\rho_{tc}$ are small because the contributions to the $H^\pm W^\mp Z$ vertices from $\rho^u$ are approximately proportional to $\sum_{l,m=2,3} (V^\dagger)_{lm} (\rho^{u \dagger} V)_{ml} f(m_{u_m}) = \rho_{cc}^* f(m_c) + \rho_{tt}^* f(m_t)$, where the function $f$ is expressed by the loop functions.
Therefore, in the following analyses, we only focus on the $\rho_{tt}$ dependence as the fermion contributions.

Third, we discuss theoretical constraints on the parameters in our model.
As the theoretical constraints for those input parameters, we take into account the Bounded From the Below (BFB) conditions~\cite{Klimenko:1984qx,Sher:1988mj,Nie:1998yn,Kanemura:1999xf,Ferreira:2004yd,Bahl:2022lio} and the perturbative unitarity~\cite{Kanemura:1993hm,Akeroyd:2000wc,Ginzburg:2005dt,Kanemura:2015ska}.
The necessary conditions for our potential to be bounded from the below are summarized in ref.~\cite{Bahl:2022lio}.
Following ref.~\cite{Kanemura:2015ska}, we employ the conditions for the perturbative unitarity in the general 2HDM.
We define $\{ a_0 \}$, which is a set of the eigenvalues of $s$-wave amplitude matrix for the elastic scatterings among the longitudinal modes of the gauge bosons and the neutral and charged scalar bosons in the high energy limit.
The criterion of $\mathrm{Max}\{ a_0 \} \equiv \xi_{\mathrm{Max}}$ is usually taken by $0.5$~\cite{Gunion:1989we} or $1$~\cite{Lee:1977eg}.

Fourth, we briefly review constraints from the high energy collider experiments and the flavor experiments.
The mixing angle $\alpha_i$ and the top Yukawa interaction $\rho_{tt}$ are constrained by the Higgs signal strengths (for available data, see table~III in ref.~\cite{Karan:2023kyj}).
As a conservative bound, we set $|\alpha_i| \le 0.01$ in the following analyses~\cite{Karan:2023kyj}.
From the combined results at the LEP~\cite{ALEPH:2013htx}, a lower bound on the mass of the charged scalar bosons is given by $m_{H^\pm} \gtrsim 80$ GeV, which is almost independent of $\rho^f$.
At the LHC, $H^\pm$ is produced via the top-associated production $gb \to H^\pm t$~\cite{LHCHiggsCrossSectionWorkingGroup:2016ypw}, and $\rho_{tt}$ is constrained by the following decay mode $H^\pm \to tb$~\cite{ATLAS:2021upq}.
The decays $H^\pm \to W^\pm Z$ via the WZ fusion production process $pp (WZ) \to H^\pm X$ have been searched by ATLAS~\cite{ATLAS:2022zuc} and CMS~\cite{CMS:2021wlt}.
These production and decay processes are the loop-induced processes in our model, and later we will evaluate the cross section by using the $H^\pm W^\mp Z$ vertices in our model.
The parameter $\rho_{tt}$ is also constrained by the direct search of the additional neutral scalar bosons $gg \to H_{2,3} \to tt$~\cite{ATLAS:2018rvc,ATLAS:2024vxm,CMS:2019pzc}.
When the mixing angles $\alpha_i$ are non-zero, the additional neutral scalar bosons $H_{2,3}$ can decay into $WW$, $ZZ$, $H_1 H_1$ or $Z H_1$, if they are kinematically allowed~\cite{Aiko:2020ksl}.
The relevant flavor observables are $B \to X_s \gamma$ and $B_s \to \mu \mu$, which are loop-induced processes in this setup~\cite{Jung:2010ab,Jung:2010ik,Jung:2012vu,Crivellin:2013wna,Li:2014fea,Crivellin:2019dun,Karan:2023kyj}.
In the following numerical results, we set a limit on $\rho_{tt}$, which satisfies the above collider and flavor constraints, as $|\rho_{tt}| \le 0.2$.

\subsection{Numerical analyses for the decay $H^+ \to W^+ Z$}
In this subsection, we give numerical results for the decay $H^+ \to W^+ Z$ in our model.

\begin{figure}[t]
  \centering
  \includegraphics[width=0.49\linewidth]{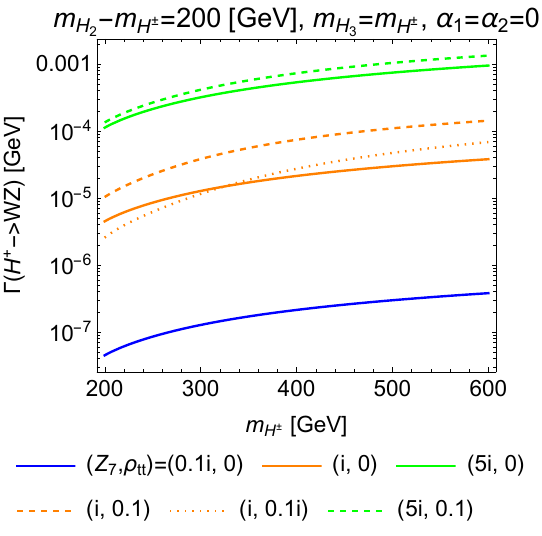}
  \includegraphics[width=0.49\linewidth]{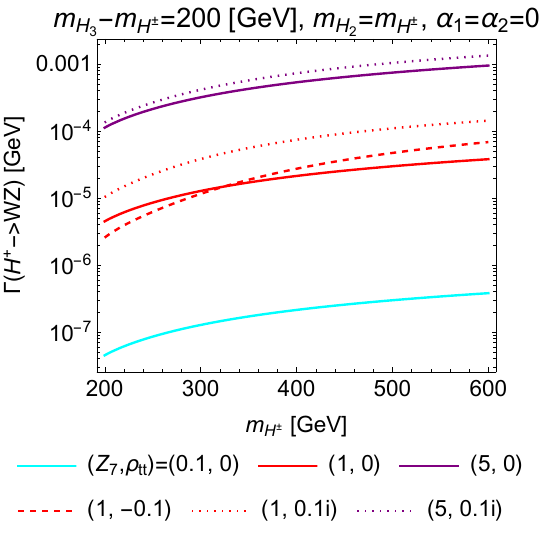}
  \caption{The decay rate $\Gamma(H^+ \to W^+ Z)$ as a function of $m_{H^\pm}$.
  In the left (right) panel, the mass difference $m_{H_2} - m_{H^\pm}$ ($m_{H_3} - m_{H^\pm}$) and the coupling constant $Z_7^I$ ($Z_7^R$) are switched on.
  The solid lines show the results for $\rho_{tt} = 0$, and both dashed and dotted lines show the impacts from non-zero $\rho_{tt}$.}
  \label{fig:decayrate_full}
\end{figure}
In figure~\ref{fig:decayrate_full}, $\Gamma(H^+ \to W^+ Z)$ is shown as a function of $m_{H^\pm}$.
In the left panel, we take $m_{H_2} - m_{H^\pm} = 200$~GeV, $m_{H_3} = m_{H^\pm}$ and $\alpha_1 = \alpha_2 = 0$.
The results of $Z_7^I = 0.1$, $1$, and $5$ are shown as the blue, orange, and green solid lines, respectively.
We note that taking $Z_7^I = 5$ is dangerous due to the perturbative unitarity or the BFB conditions, however, it is shown just for illustration.
These solid lines are purely caused by the scalar and gauge contributions, and the decay rate increases as taking large $Z_7^I$.
This is because, in the left panel, the custodial symmetry violation in the potential is realized by non-zero $Z_7^I$.
When the fermion contribution from $\rho_{tt}$ is switched on, as shown by the dashed ($\rho_{tt} = 0.1$) and the dotted ($\rho_{tt} = 0.1i$) lines, the decay width is changed.
In the right panel of figure~\ref{fig:decayrate_full}, $m_{H_3} - m_{H^\pm} = 200$ GeV and $m_{H_2} = m_{H^\pm}$ are taken.
Namely, the masses of $H_2$ and $H_3$ are exchanged from the left panel.
The cyan, red, and purple lines are the results of $Z_7^R = 0.1$, $1$, and $5$, respectively, and the impacts of setting $\rho_{tt} = -0.1$ and $0.1 i$ are shown by the dashed and dotted lines, respectively.
In this right panel, the violation of the twisted custodial symmetry in the potential is realized by $Z_7^R \neq 0$.
We can see the equivalent results in the left and right panels, which are related to the violation of the custodial or the twisted custodial symmetry.

\begin{figure}[t]
  \centering
  \includegraphics[width=0.46\linewidth]{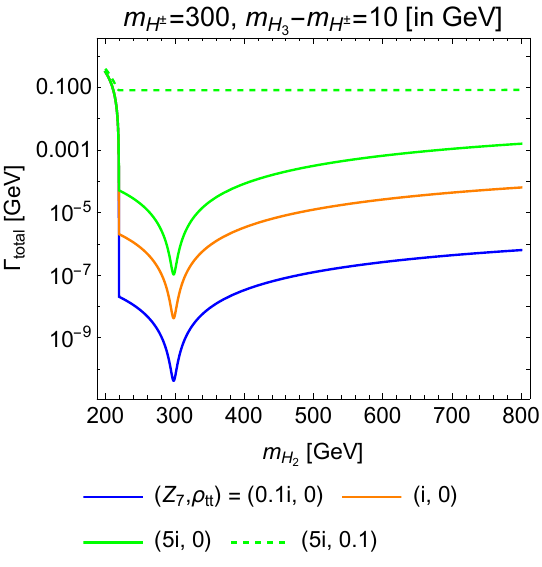}
  \includegraphics[width=0.49\linewidth]{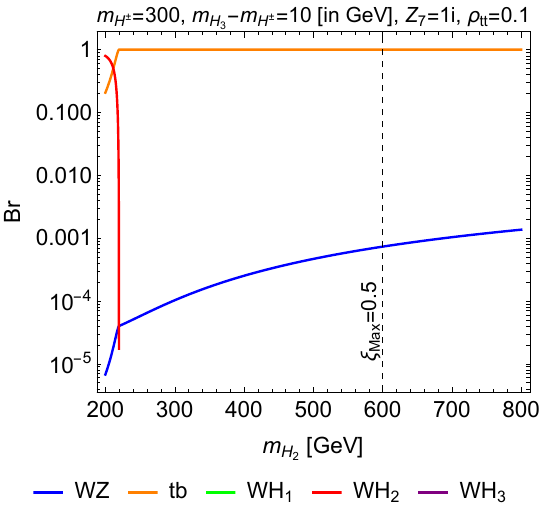}
  \caption{
  Left: $m_{H_2}$ dependence of the total decay rate of $H^+$.
  The blue, orange, and green lines show the results of $Z_7^I = 0.1$, $1$, and $5$, respectively.
  The result with non-zero $\rho_{tt}$ is shown as the dashed line.
  Right: $m_{H_2}$ dependence of the branching ratio of $H^+$.
  The decay modes are $H^+ \to W^+ Z$ (blue), $tb$ (orange), $W^+ H_1$ (green), $W^+ H_2$ (red), and $W^+ H_3$ (purple).
  In this parameter set, only the decay modes $H^+ \to W^+ Z$, $tb$ and $W^+ H_2$ can be significant.
  The black dashed line shows the criterion where the maximal $s$-wave scattering amplitude among the scalar and the gauge bosons is 0.5.}
  \label{fig:WidthandBR2}
\end{figure}
In the left panel of figure~\ref{fig:WidthandBR2}, $m_{H_2}$ dependence of the total decay width of $H^+$ is shown.
We have taken $m_{H^\pm} = 300$~GeV, $m_{H_3} - m_{H^\pm} = 10$~GeV and $\alpha_1 = \alpha_2 = 0$.
In the left panel, the blue, orange, and green solid lines show the total decay width $\Gamma_{\mathrm{total}}$ with $Z_7^I = 0.1$, $1$, and $5$.
Again, taking $Z_7^I = 5$ is dangerous due to the perturbative unitarity or the BFB conditions, however, the result is shown just for illustration.
When $m_{H_2} < m_{H^\pm} - m_W$, the $H^+ \to W^+ H_2$ decay is kinematically allowed.
On the other hand, in the region $m_{H_2} > m_{H^\pm} - m_W$, only the decay mode $H^+ \to W^+ Z$ is possible if $\rho_{tt} = 0$, and it almost disappears at the point $m_{H^\pm} = m_{H_2}$, where the potential recovers the twisted custodial symmetry.
As shown in the green dashed line, when $\rho_{tt}$ is switched on, $H^+ \to tb$ decay is allowed.
In the right panel of figure~\ref{fig:WidthandBR2}, the branching ratio of $H^+$ is shown.
In this panel, $Z_7^I = 1$ and $\rho_{tt} = 0.1$ are taken.
The decay modes are $H^+ \to W^+ Z$ (blue), $tb$ (orange), $W^+ H_1$ (green), $W^+ H_2$ (red), and $W^+ H_3$ (purple). 
In this parameter set, only $H^+ \to W^+ Z$, $tb$ and $W^+ H_2$ can be significant.
The black dashed line shows the perturbative unitarity bound where $\xi_{\mathrm{Max}}$ is 0.5.
The line for $\xi_{\mathrm{Max}} = 1$ is out of this figure.
In the region $m_{H_2} > m_{H^\pm} - m_W$, the dominant decay modes are $H^+ \to tb$ and $H^+ \to W^+ Z$.
As the mass difference $m_{H_2} - m_{H^\pm}$ becomes large, the branching ratio $\mathrm{Br}(H^+ \to W^+ Z)$ increases.

\begin{figure}[t]
  \centering
  \includegraphics[width=0.49\linewidth]{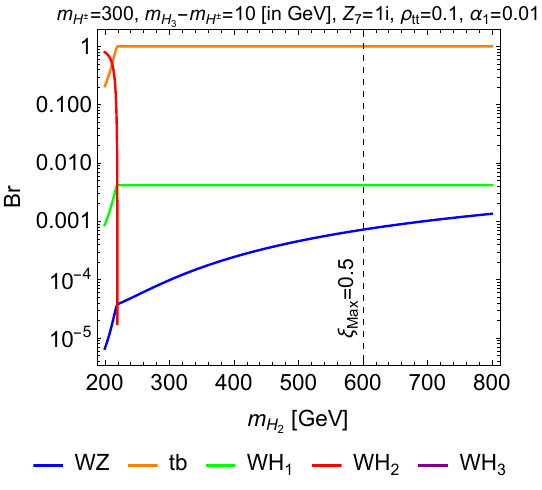}
  \includegraphics[width=0.5\linewidth]{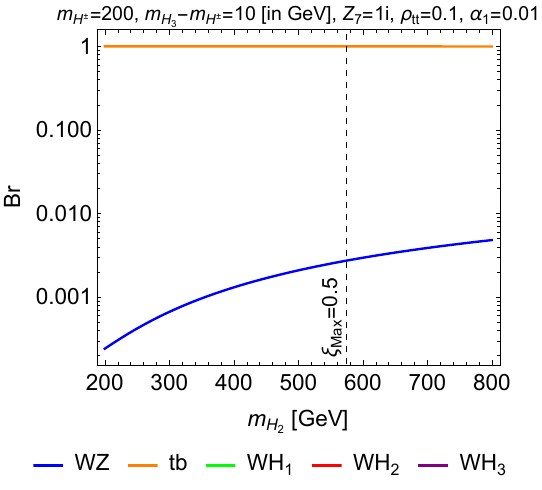}
  \caption{$m_{H_2}$ dependence of the branching ratio of $H^+$ with the non-zero mixing angle $\alpha_1$ (left: $m_{H^\pm} = 300$~GeV, right: $m_{H^\pm} = 200$~GeV).}
  \label{fig:BRnonalign}
\end{figure}
In figure~\ref{fig:BRnonalign}, analyses for the branching ratio with a non-alignment case are shown.
In the left (right) panel, $m_{H^\pm} = 300$~GeV ($m_{H^\pm} = 200$~GeV), $m_{H_3} = m_{H^\pm} + 10$~GeV, $Z_7^I = 1$, $\rho_{tt} = 0.1$, $\alpha_1 = 0.01$, and $\alpha_2 = 0$ are taken.
As shown in the left panel, when $m_{H^\pm} > m_W + m_{H_1}$, $\mathrm{Br}(H^+ \to W^+ Z)$ is suppressed by the $H^+ \to W^+ H_1$ decay, which is caused by non-zero $\mathcal{R}_{12}$.
On the other hand, when $m_{H^\pm} < m_W + m_{H_1}$, as shown in the right panel, $H^+$ decays into $tb$ or $W^+ Z$, so that $\mathrm{Br}(H^+ \to W^+ Z) \gtrsim O(10^{-3})$ is realized for $\xi_{\mathrm{Max}} \le 0.5$.

\begin{figure}[t]
  \centering
  \includegraphics[width=0.49\linewidth]{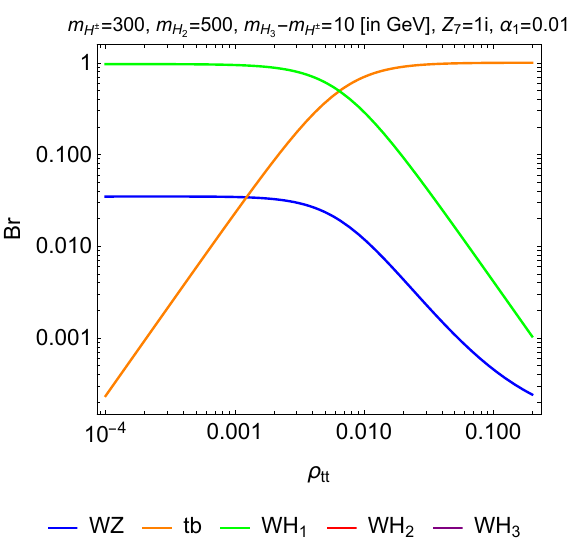}
  \includegraphics[width=0.5\linewidth]{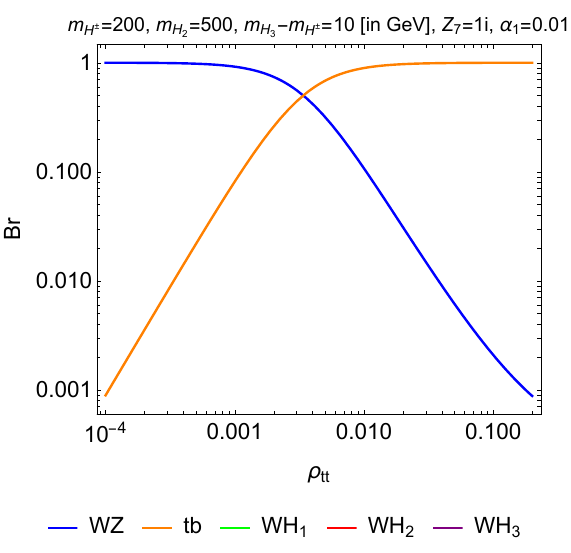}
  \caption{$\rho_{tt}$ dependence of the branching ratio of $H^+$ (left: $m_{H^\pm} = 300$~GeV, right: $m_{H^\pm} = 200$~GeV).
  In this parameter set, the decay modes $H^+ \to W^+ Z$, $tb$ and $W^+ H_1$ are relevant.}
  \label{fig:Rhott}
\end{figure}
In figure~\ref{fig:Rhott}, the results for $\rho_{tt}$ dependence of the branching ratio are shown.
As shown in the left panel, where $m_{H^\pm} = 300$~GeV, the dominant decay mode is $H^+ \to W^+ H_1$ in the region of $\rho_{tt} \lesssim 0.006$, and $H^+ \to tb$ becomes dominant as $\rho_{tt}$ increases.
In the right panel, where $m_{H^\pm} = 200$~GeV, $H^+ \to W^+ H_1$ is kinematically forbidden.
As a result, in the region of $\rho_{tt} \lesssim 0.003$, $H^+ \to tb$ is suppressed, and $\mathrm{Br}(H^+ \to W^+ Z)$ can be large.

\subsection{Asymmetry in the decays of $H^+ \to W^+ Z$ and $H^- \to W^- Z$}

In the above section, the numerical results for $H^+$ decays are shown.
The decay $H^- \to W^- Z$ also behaves in a similar way to the decay $H^+ \to W^+ Z$.
However, an asymmetry between them is caused by the interference of the scalar and fermion contributions.
In this subsection, we give numerical results for the CP asymmetry in these decays. 

For the discussions of the CP violation, we define 
\begin{align}
  \Delta (H^\pm \to W^\pm Z) \equiv \Gamma(H^+ \to W^+ Z) - \Gamma(H^- \to W^- Z),
\end{align}
and the CP violating quantity~\cite{Arhrib:2007rm}
\begin{align}
  \delta_{\mathrm{CP}} \equiv \frac{\Gamma(H^+ \to W^+ Z) - \Gamma(H^- \to W^- Z)}{\Gamma(H^+ \to W^+ Z) + \Gamma(H^- \to W^- Z)}.
  \label{eq:deltaCP}
\end{align}
By definition, the decay rate for $H^-$ is given by 
\begin{align}
  \Gamma(H^- \to W^- Z) = \frac{1 - \delta_{\mathrm{CP}}}{1 + \delta_{\mathrm{CP}}} \Gamma(H^+ \to W^+ Z).
  \label{eq:Hminusdecay}
\end{align}

\begin{figure}[t]
  \centering
  \includegraphics[width=1.0\linewidth]{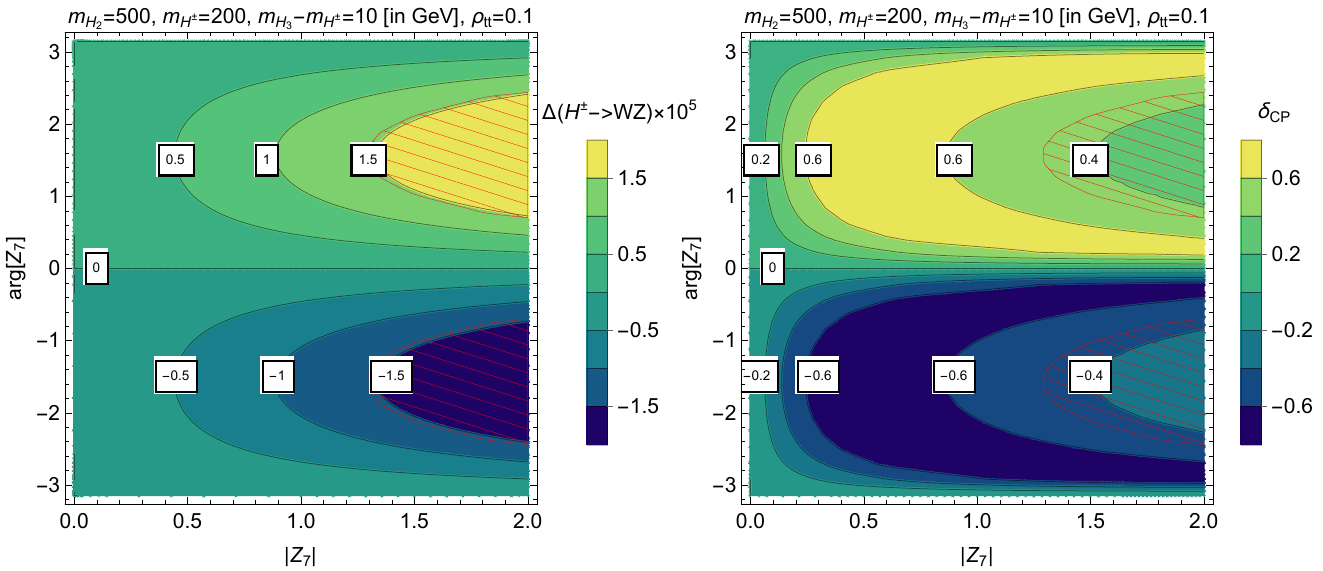}
  \caption{The contour plots of $\Delta(H^\pm \to W^\pm Z) \times 10^5$ (left) and $\delta_{\mathrm{CP}}$ (right) in $|Z_7|$-$\mathrm{arg}[Z_7]$ plane.
  The red shaded regions do not satisfy the BFB conditions.}
  \label{fig:ContourCPV}
\end{figure}
The behavior of these quantities are shown in figure~\ref{fig:ContourCPV}, where $m_{H^\pm} = 200$~GeV, $m_{H_2} = 500$~GeV, $m_{H_3} = m_{H^\pm} + 10$~GeV, $\rho_{tt} = 0.1$, and $\alpha_1 = \alpha_2 = 0$ are taken.
In the left and right panel, the contour figures of $\Delta (H^\pm \to W^\pm Z) \times 10^{5}$ and $\delta_{CP}$ in the $|Z_7|$-$\mathrm{arg}[Z_7]$ plane are shown, respectively.
The red shaded regions do not satisfy the BFB conditions.
At $\mathrm{arg}[Z_7] = \pi/2$, $\Delta (H^\pm \to W^\pm Z)$ and $\delta_{\mathrm{CP}}$ take the maximal value, while at $\mathrm{arg}[Z_7] = -\pi/2$, they take the minimal value.
This feature can be understood as follows: each decay amplitude can be approximately written as 
\begin{align}
  &\mathcal{M}(H^+ \to W^+ Z) \simeq i \big(\rho_{tt}^R f_1 + Z_7^R (m_{H^\pm}^2 - m_{H_3}^2) f_2 \big) + \big( \rho_{tt}^I f_1 + Z_7^I  (m_{H^\pm}^2 - m_{H_2}^2) f_3 \big), \notag \\
  &\mathcal{M}(H^- \to W^- Z) \simeq -i \big(\rho_{tt}^R f_1 + Z_7^R (m_{H^\pm}^2 - m_{H_3}^2) f_2 \big) + \big(\rho_{tt}^I f_1 + Z_7^I (m_{H^\pm}^2 - m_{H_2}^2) f_3 \big),
\end{align}
where $f_{1,2,3}$ are mass dependent functions.
When $m_{H^\pm} > m_t + m_b$, the loop functions in $f_1$ have the imaginary parts, so that $\delta_{\mathrm{CP}}$ can be expressed as  
\begin{align}
  &\delta_{\mathrm{CP}} \propto \Delta (H^\pm \to W^\pm Z) \propto |\mathcal{M}(H^+ \to W^+ Z)|^2 - |\mathcal{M}(H^- \to W^- Z)|^2 \notag \\
  &\propto \rho_{tt}^R Z_7^I (m_{H^\pm}^2 - m_{H_2}^2) f_3 \mathrm{Im}[f_1^*] + \rho_{tt}^I Z_7^R (m_{H^\pm}^2 - m_{H_3}^2) f_2 \mathrm{Im}[f_1].
\end{align}
As a result, the dependence of $Z_7^I \propto \sin ( \mathrm{arg}[Z_7])$ appears in figure~\ref{fig:ContourCPV}, where non-zero $m_{H^\pm} - m_{H_2}$ and $\rho_{tt}^R$ are taken.
Due to the denominator in eq.~(\ref{eq:deltaCP}), as $|Z_7|$ increases, $\delta_{\mathrm{CP}}$ gets large at first, but it turns to decrease. 

\begin{figure}[t]
  \centering
  \includegraphics[width=0.6\linewidth]{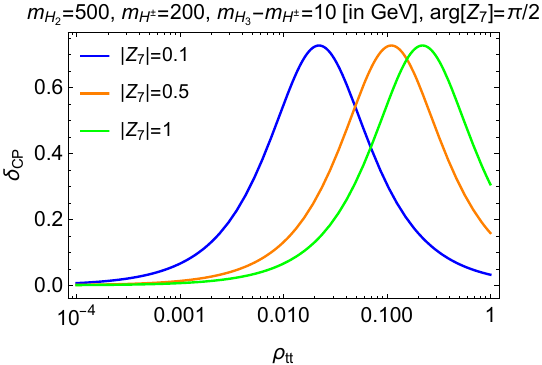}
  \caption{The $\rho_{tt}$ dependence of $\delta_{\mathrm{CP}}$ with several $|Z_7|$ values.
  $|Z_7| = 0.1$ (blue), $|Z_7| = 0.5$ (orange), and $|Z_7| = 1$ (green).}
  \label{fig:ContourCPV3}
\end{figure}
In figure~\ref{fig:ContourCPV3}, the $\rho_{tt}$ dependence of $\delta_{\mathrm{CP}}$ is shown.
We have set the phase of $Z_7$ as $\mathrm{arg}[Z_7] = \pi/2$. 
The blue, orange, and green lines show the results of $|Z_7| = 0.1$, $0.5$ and $1$, respectively.
The point of $\rho_{tt}$, where $\delta_{\mathrm{CP}}$ takes a maximal value, depends on the size of $Z_7$.
The maximal value of $\delta_{\mathrm{CP}}$ in this figure is around 0.7, so that $\Gamma(H^- \to W^- Z)$ is about $17~\%$ of $\Gamma(H^+ \to W^+ Z)$.

At the $pp$ collider like LHC, production cross sections of $H^+$ and $H^-$ via the vector boson fusion mechanism and the associated production mechanism with neutral scalar bosons are in general different~\cite{Kanemura:2001hz,Cao:2003tr,Asakawa:2005gv}.
In addition, as we find, the decays of $H^+ \to W^+ Z$ and $H^- \to W^- Z$ can be different by the CP violating phase effect.

\section{Discussions \label{sec:discussion}}
\begin{figure}[t]
  \centering
  \includegraphics[width=0.5\linewidth]{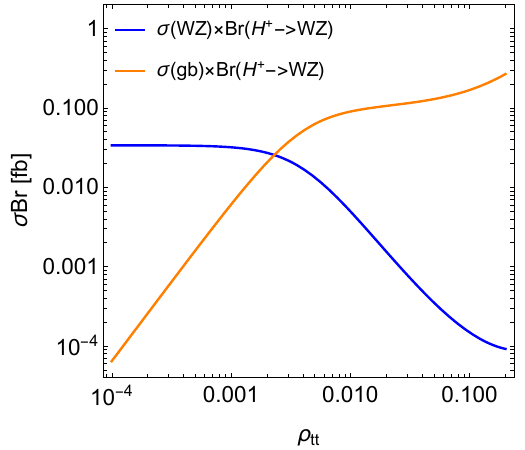}
  \caption{$\rho_{tt}$ dependence of the cross section $\sigma(pp \to H^+ X \to W^+ Z X)$ in fb.
  The charged scalar boson is produced via the WZ fusion process (blue) or the top-associated process (orange).}
  \label{fig:sigmaBR}
\end{figure}

First, we discuss the signature for the $H^\pm W^\mp Z$ vertices at the hadron colliders.
Based on the analyses in the former sections, we set two benchmark points which are given in table~\ref{tab:BP}.
The difference in BP1 and BP2 is the size of $\rho_{tt}$, and the other parameters are the same.
\begin{table}[t]
  \centering
  \begin{tabular}{|c|c|c|c|c|c|c|c|}
      \hline
      (in GeV) & $m_{H^\pm}$ & $m_{H_2}$ & $m_{H_3}$ & $Z_7$ & $\rho_{tt}$ & $\alpha_1 = - \alpha_2$ \\ \hline
  BP1 &  200    &  500   &  210   &  $1.3 e^{2.0 i}$  &   0.1    &   0.01       \\ \hline
  BP2 &  200    &  500   & 210    &  $1.3 e^{2.0 i}$  &   0.001    &   0.01    \\ \hline 
  \end{tabular}
  \caption{Benchmark points for the discussions of collider phenomenology.
  In BP1 and BP2, the same parameters are taken except for $\rho_{tt}$.}
  \label{tab:BP}
\end{table}
As discussed in section~\ref{sec:vertex}, for the production of $H^+$, the WZ fusion process ($pp(W^+ Z) \to H^+ X$) and the top-associated process ($gb \to H^+ t$) are important in the hadron collider.
(At a $pp$ collider, the production cross section for $pp(W^- Z) \to H^- X$ is smaller than that for $pp(W^+ Z) \to H^+ X$~\cite{Asakawa:2005gv}.)
The former production process is loop-induced, and it has been evaluated in refs.~\cite{Asakawa:2005gv,Asakawa:2006gm,Abbas:2018pfp}.
From figure~13 in ref.~\cite{Abbas:2018pfp}, the WZ fusion cross section can be obtained as $\sigma_{WZ} \simeq 2 |F|^2 \times 10^{3}$ fb for $m_{H^\pm} = 200$~GeV at the LHC with $\sqrt{s} = 13$~TeV.
The latter process depends on $\rho_{tt}$, and its cross section has been evaluated by $\sigma_{gb} \simeq 10^2$ fb for $|\rho_{tt}| \simeq 1/8$ and $m_{H^\pm} = 200$~GeV at the LHC with $\sqrt{s} = 13$~TeV~\cite{LHCHiggsCrossSectionWorkingGroup:2016ypw}.
Therefore, $\rho_{tt}$ dependence of this cross section can be estimated as $\sigma_{gb} \simeq 64 |\rho_{tt}|^2 \times 10^2$ fb for $m_{H^\pm} = 200$ GeV.
In figure~\ref{fig:sigmaBR}, $\rho_{tt}$ dependence of $\sigma_{WZ} \times \mathrm{Br}(H^+ \to W^+ Z)$ and $\sigma_{gb} \times \mathrm{Br}(H^+ \to W^+ Z)$ are shown.
The point $\rho_{tt} = 0.1$ ($0.001$) in figure~\ref{fig:sigmaBR} corresponds to BP1 (BP2).
The numerical values of $|F|^2$, $\mathrm{Br}(H^+ \to W^+ Z)$, and the cross sections at these points are summarized in table~\ref{tab:summarize}.
For relatively large $\rho_{tt}$, e.g. in BP1, in spite of the suppression of $\mathrm{Br}(H^+ \to W^+ Z)$, $\sigma_{gb}\times \mathrm{Br}$ has advantage because the top-associated production process is enhanced by $\rho_{tt}$.
On the other hand, in the case of relatively small $\rho_{tt}$, e.g. in BP2, the top-associated production process and the $H^+ \to tb$ decay are suppressed.
As a result, the WZ fusion production process is important in this case.
\begin{table}[t]
  \centering
  \begin{tabular}{|c|c|c|}
      \hline
       & BP1 & BP2 \\ \hline \hline
  $|F|^2$ &  $2.8 \times 10^{-5}$    &  $1.7 \times 10^{-5}$   \\ \hline
  $\mathrm{Br}(H^+ \to W^+ Z)$ &  $2.6 \times 10^{-3}$    &  $9.4 \times 10^{-1}$   \\ \hline
  $\sigma_{WZ}$ [fb] &  $5.6 \times 10^{-2}$   &  $3.4 \times 10^{-2}$ \\ 
  $\sigma_{WZ} \times \mathrm{Br}$ [fb] &  $1.5 \times 10^{-4}$   &  $3.2 \times 10^{-2}$  \\ \hline 
  $\sigma_{gb}$ [fb] & $6.4 \times 10$ &  $6.4 \times 10^{-3}$  \\ 
  $\sigma_{gb} \times \mathrm{Br}$ [fb] &  $1.7 \times 10^{-1}$  &  $6.0 \times 10^{-3}$  \\ \hline
  \end{tabular}
  \caption{Summary of several values for $|F|^2$, $\mathrm{Br}(H^+ \to W^+ Z)$ and the cross sections in BP1 and BP2.}
  \label{tab:summarize}
\end{table}

From the data collected at the 13~TeV LHC with 139~$\mathrm{fb}^{-1}$~\cite{ATLAS:2022zuc}, an upper bound $\sigma_{WZ} \times \mathrm{Br} \lesssim O(10^{2})$~fb for $m_{H^\pm} = 200$~GeV has been obtained, which is much larger than the estimated value in BP2.
However, at the HL-LHC with $3000~\mathrm{fb}^{-1}$, we can expect $O(10^2)$ number of this events in BP2.
This signal is also important in the high energy upgrade of the LHC~\cite{Adhikary:2020cli}.
Even in BP1, in which $\rho_{tt}$ is relatively large, the process $\sigma_{gb} \times \mathrm{Br}$ might be important in the future hadron colliders because $\sigma_{gb} \times \mathrm{Br}$ in BP1 is about 5 times larger than $\sigma_{WZ} \times \mathrm{Br}$ in BP2.
Consequently, the $H^\pm W^\mp Z$ vertices in our model are expected to be tested at the HL-LHC.

Second, we give discussions for the testability in the lepton collider.
For example, in the high energy $e^+ e^-$ collider such as the ILC with $\sqrt{s} = 500$ GeV and $1$ TeV~\cite{Fujii:2015jha,Bambade:2019fyw}, the pair production process $e^+ e^- \to H^+ H^-$ is important, and the cross section only depends on $m_{H^\pm}$~\cite{Komamiya:1988rs}.
In addition, single $H^\pm$ production has been discussed in refs.~\cite{Gunion:1988tf,Kanemura:2000cw,Kanemura:2011kc}.
When we consider $m_{H^\pm} = 200$~GeV, the production cross section of a pair of the charged scalar bosons is given by $\sigma(H^+H^-) \simeq 3.0 \times 10$~fb ($2.0 \times 10$~fb) at the ILC $500$~GeV ($1$~TeV)~\cite{Kanemura:2000cw}.
Therefore, in BP1 (BP2), $\sigma(H^+ H^-) \times \mathrm{Br}(H^+ \to W^+ Z)$ is estimated as $\sigma \times \mathrm{Br} \simeq 7.8 \times 10^{-2}$~fb ($2.8 \times 10$~fb) at the ILC $500$~GeV and $\sigma \times \mathrm{Br} \simeq 5.2 \times 10^{-2}$~fb ($1.9 \times 10$~fb) at the ILC $1$~TeV.
This pair production process is independent of $\rho_{tt}$, so that numerous events can be expected in BP2, in which almost all $H^+$ decays to the W and Z bosons.
More detailed analysis in these hadron and lepton colliders will be given in the following paper.

Third, our calculation has been done in the Higgs basis, and it can be applied to the softly-broken $\mathbb{Z}_2$ symmetric 2HDM by using the basis rotation from the $\mathbb{Z}_2$ basis to the Higgs basis~\cite{Davidson:2005cw}.
As shown in table~\ref{tab:Z2}, the CP violating phases do not exist in the Yukawa sector in the softly-broken $\mathbb{Z}_2$ symmetric 2HDM, and the $\rho^f$ coupling constants are not independent of the parameters in the potential.
On the other hand, in the general 2HDM, $\rho^f$ are the arbitrary complex matrices, and they can have the CP violating phases, some of which are important sources to produce the BAU via the mechanism of the electroweak baryogenesis~\cite{Cline:2011mm,Tulin:2011wi,Liu:2011jh,Ahmadvand:2013sna,Chiang:2016vgf,Guo:2016ixx,Fuyuto:2017ewj,Modak:2018csw,Enomoto:2021dkl,Enomoto:2022rrl,Kanemura:2023juv}.
Due to the independence between $\rho^f$ and the coupling constants in the potential, it is possible to consider parameter points, such as BP2, in which $\mathrm{Br}(H^+ \to W^+ Z)$ is close to unity.

Finally, in this paper, we have discussed the relation between the $H^\pm W^\mp Z$ vertices and the CP violation.
As we have mentioned in section~\ref{sec:symmetry}, the decays $H^\pm \to W^\pm Z$ cannot be the observable for $\mathrm{Im}[Z_5^* Z_7^2]$, which is the CP violating invariant in the potential.
However, the difference in the decays $H^+ \to W^+ Z$ and $H^- \to W^- Z$ is sensitive to $\mathrm{Im}[Z_7 \rho^f]$, as we have discussed in section~\ref{sec:vertex}.
Therefore, as shown in figure~\ref{fig:ContourCPV}, if we know the phase of $\rho^f$, the phase of $Z_7$ can be determined by measuring the difference between $\Gamma(H^+ \to W^+ Z)$ and $\Gamma(H^- \to W^- Z)$.
For example, in our model, it has been known that the rephasing invariant $Z_5 \rho_{\tau \tau}^2$ can be measured by using azimuthal angle dependence of the hadronic decay of the tau leptons, which are decay products of the pair of $H_2$ and $H_3$~\cite{Kanemura:2020ibp,Kanemura:2021atq}.
More realistic analyses for measuring the CP phase via the $H^\pm W^\mp Z$ vertices will be discussed somewhere~\cite{Kanemura:FutureWork}.

\section{Conclusion}

In this paper, we have discussed the $H^\pm W^\mp Z$ vertices in the most general 2HDM with the CP violation.
It has been known that the CP violating potential in the 2HDM violates the custodial symmetry.
The $H^\pm W^\mp Z$ vertices are one of the phenomenological consequences for the custodial symmetry violation.
We have calculated the $H^\pm W^\mp Z$ vertices at the one-loop level with the most general setup, and we have found that there are both contributions from the CP conserving and CP violating coupling constants.
The scalar loop contributions disappear at the one-loop level when the custodial symmetry is respected in the Higgs potential.
We have discussed the decays $H^\pm \to W^\pm Z$ in several parameter sets.
We have found that the decay rates can be significantly enhanced by the loop effects of the additional scalar bosons in the most general 2HDM with the CP violation.
We have also found that the asymmetry between the decay rates of $H^+ \to W^+ Z$ and $H^- \to W^- Z$ can be an important observable to see the CP violating phases in the model. 
Finally, we have given discussions for the testability of these vertices at current and future high energy collider experiments.

\section*{Acknowledgments}
The work of S.~K. was supported by the JSPS Grants-in-Aid for Scientific Research No.~20H00160 and No.~23K17691.
The work of Y.~M. was supported by the JSPS Grant-in-Aid for JSPS Fellows No.~23KJ1460.

\newpage

\appendix

\section{Definition of the loop functions \label{sec:app2}}
We here show the definitions of the Passarino--Veltmann loop functions.
\begin{align}
  &\frac{1}{16\pi^2 } A_0 [m^2] =  \mu^{2\epsilon} \int \frac{ d^D l}{i (2\pi)^D} \frac{1}{l^2 - m^2 + i\varepsilon}, \notag \\
  &\frac{1}{16\pi^2 } B_{\{0,\mu,\mu\nu\}} [k; m_a^2, m_b^2] = \mu^{2\epsilon} \int \frac{d^D l}{i (2\pi)^D} \frac{\{1, l_\mu, l_\mu l_\nu\}}{l^2 - m_a^2 + i\varepsilon}\frac{1}{(l+k)^2 - m_b^2 + i\varepsilon},\notag \\
  &\frac{1}{16\pi^2 } C_{\{ 0,\mu,\mu \nu \}} [k_1, k_2; m_a^2,m_b^2,m_c^2] =\notag \\
  &\qquad \mu^{2\epsilon} \int \frac{d^D l}{i (2\pi)^D} \frac{\{1, l_\mu, l_\mu l_\nu\}}{l^2 - m_a^2 + i\varepsilon}\frac{1}{(l+k_1)^2 - m_b^2 + i\varepsilon} \frac{1}{(l+k_1 + k_2)^2 - m_c^2 + i\varepsilon},
\end{align}
where $D = 4 - 2 \epsilon$.
The tensor decompositions are performed as 
\begin{align}
  &B_{\mu}[k; m_a^2, m_b^2] = B_1 k_\mu, \notag \\
  &B_{\mu \nu}[k; m_a^2, m_b^2] = B_{21} k_\mu k_\nu + B_{22} g_{\mu \nu}, \notag \\
  &C_{\mu}[k_1, k_2; m_a^2,m_b^2,m_c^2] = C_{11} k_{1 \mu} + C_{12} k_{2 \mu}, \notag \\
  &C_{\mu \nu}[k_1, k_2; m_a^2,m_b^2,m_c^2] = C_{21} k_{1 \mu}k_{1 \nu} + C_{22} k_{2 \mu} k_{2 \nu} + C_{23} (k_{1\mu} k_{2 \nu} + k_{2 \mu} k_{1 \nu}) + C_{24} g_{\mu\nu}.
\end{align}

\section{Analytic formulae for the $H^\pm W^\mp Z$ vertices in the general two Higgs doublet model \label{sec:app1}}

In this appendix, we give analytic formulae for the diagrams which are shown in figures~\ref{fig:HWZ_full_Ctype}-\ref{fig:HWZ_A}.

First, the $C$ type contributions are given by 
\begin{align}
  &F^{(C1)} _{+}= \frac{-2i}{16 \pi^2 c_W} (\mathcal{R}_{i2} - i\mathcal{R}_{i3}) (\mathcal{R}_{i3} \mathcal{R}_{j2} - \mathcal{R}_{i2} \mathcal{R}_{j3}) (Z_3 \mathcal{R}_{j1} + Z_7^R \mathcal{R}_{j2} - Z_7^I \mathcal{R}_{j3}) C_{24}[m_{H^\pm}^2, m_{H_i}^2, m_{H_j}^2], \notag \\
  &G^{(C1)} _{+}= \frac{-2i m_W^2}{16 \pi^2 c_W} (\mathcal{R}_{i2} - i\mathcal{R}_{i3}) (\mathcal{R}_{i3} \mathcal{R}_{j2} - \mathcal{R}_{i2} \mathcal{R}_{j3}) (Z_3 \mathcal{R}_{j1} + Z_7^R \mathcal{R}_{j2} - Z_7^I \mathcal{R}_{j3}) (C_{23} + C_{12})[m_{H^\pm}^2, m_{H_i}^2, m_{H_j}^2], \notag \\
  &F^{(C2)} _{+}= \frac{2 c_{2W}}{16 \pi^2 c_W} (\mathcal{R}_{i2} - i\mathcal{R}_{i3}) (Z_3 \mathcal{R}_{i1} + Z_7^R \mathcal{R}_{i2} - Z_7^I \mathcal{R}_{i3}) C_{24}[m_{H_i}^2, m_{H^\pm}^2, m_{H^\pm}^2], \notag \\
  &G^{(C2)} _{+}= \frac{2 m_W^2 c_{2W}}{16 \pi^2 c_W} (\mathcal{R}_{i2} - i\mathcal{R}_{i3}) (Z_3 \mathcal{R}_{i1} + Z_7^R \mathcal{R}_{i2} - Z_7^I \mathcal{R}_{i3}) (C_{23} + C_{12})[m_{H_i}^2, m_{H^\pm}^2, m_{H^\pm}^2], \notag \\
  &F^{(C3)} _{+}= \frac{-2i }{16 \pi^2 c_W} \mathcal{R}_{i1} (\mathcal{R}_{i3} \mathcal{R}_{j2} - \mathcal{R}_{i2} \mathcal{R}_{j3}) \Big(\frac{1}{2}(Z_4 + Z_5) \mathcal{R}_{j2} - \frac{i}{2} (Z_4 - Z_5) \mathcal{R}_{j3} + Z_6 \mathcal{R}_{j1} \Big) C_{24}[m_{G^\pm}^2, m_{H_i}^2, m_{H_j}^2], \notag \\
  &G^{(C3)} _{+}= \frac{-2i m_W^2}{16 \pi^2 c_W} \mathcal{R}_{i1} (\mathcal{R}_{i3} \mathcal{R}_{j2} - \mathcal{R}_{i2} \mathcal{R}_{j3}) \notag \\
  &\qquad \qquad \times \Big(\frac{1}{2}(Z_4 + Z_5) \mathcal{R}_{j2} - \frac{i}{2} (Z_4 - Z_5) \mathcal{R}_{j3} + Z_6 \mathcal{R}_{j1} \Big) (C_{23}+C_{12})[m_{G^\pm}^2, m_{H_i}^2, m_{H_j}^2], \notag \\
  &F^{(C4)} _{+}= \frac{-2 }{16 \pi^2 c_W} \mathcal{R}_{i1} \Big(\frac{1}{2}(Z_4 + Z_5) \mathcal{R}_{i2} - \frac{i}{2} (Z_4 - Z_5) \mathcal{R}_{i3} + Z_6 \mathcal{R}_{i1} \Big) C_{24}[ m_{G^\pm}^2, m_{G^0}^2, m_{H_i}^2], \notag \\
  &G^{(C4)} _{+}= \frac{-2 m_W^2}{16 \pi^2 c_W} \mathcal{R}_{i1} \Big(\frac{1}{2}(Z_4 + Z_5) \mathcal{R}_{i2} - \frac{i}{2} (Z_4 - Z_5) \mathcal{R}_{i3} + Z_6 \mathcal{R}_{i1} \Big) (C_{23} + C_{12})[ m_{G^\pm}^2, m_{G^0}^2, m_{H_i}^2], \notag \\
  &F^{(C5)} _{+}= \frac{2 c_{2W}}{16 \pi^2 c_W} \mathcal{R}_{i1} \Big(\frac{1}{2}(Z_4 + Z_5) \mathcal{R}_{i2} - \frac{i}{2} (Z_4 - Z_5) \mathcal{R}_{i3} + Z_6 \mathcal{R}_{i1} \Big) C_{24}[ m_{H_i}^2, m_{G^\pm}^2, m_{G^\pm}^2], \notag \\
  &G^{(C5)} _{+}= \frac{2 m_W^2 c_{2W}}{16 \pi^2 c_W} \mathcal{R}_{i1} \Big(\frac{1}{2}(Z_4 + Z_5) \mathcal{R}_{i2} - \frac{i}{2} (Z_4 - Z_5) \mathcal{R}_{i3} + Z_6 \mathcal{R}_{i1} \Big) (C_{23} + C_{12})[ m_{H_i}^2, m_{G^\pm}^2, m_{G^\pm}^2], \notag \\
  &F^{(C6)} _{+}= \frac{2i}{16 \pi^2 c_W} \frac{m_W^2}{v^2} \mathcal{R}_{i1} (\mathcal{R}_{i3}\mathcal{R}_{j2}-\mathcal{R}_{i2}\mathcal{R}_{j3})(\mathcal{R}_{j2}-i\mathcal{R}_{j3}) C_{24}[ m_{W}^2, m_{H_i}^2, m_{H_j}^2], \notag \\
  &G^{(C6)} _{+}= \frac{2im_W^2}{16 \pi^2 c_W} \frac{m_W^2}{v^2} \mathcal{R}_{i1} (\mathcal{R}_{i3}\mathcal{R}_{j2}-\mathcal{R}_{i2}\mathcal{R}_{j3})(\mathcal{R}_{j2}-i\mathcal{R}_{j3}) (C_{23}+ C_{12} + 2 C_{11} + 2 C_0)[ m_{W}^2, m_{H_i}^2, m_{H_j}^2], \notag \\
  &F^{(C7)} _{+}=  \frac{2 c_{2W}}{16 \pi^2 c_W} \frac{m_Z^2}{v^2} \mathcal{R}_{i1} (\mathcal{R}_{i2}-i\mathcal{R}_{i3}) C_{24}[ m_{H^\pm}^2, m_{H_i}^2, m_Z^2], \notag \\
  &G^{(C7)} _{+}= \frac{2 m_W^2 c_{2W}}{16 \pi^2 c_W} \frac{m_Z^2}{v^2} \mathcal{R}_{i1} (\mathcal{R}_{i2}-i\mathcal{R}_{i3}) (C_{23} - C_{12})[ m_{H^\pm}^2, m_{H_i}^2, m_Z^2], \notag \\
  &F^{(C8)} _{+}= \frac{2s_W^2 m_Z^2}{16 \pi^2 c_W} \mathcal{R}_{i1} \Big(\frac{1}{2}(Z_4 + Z_5) \mathcal{R}_{i2} - \frac{i}{2} (Z_4 - Z_5) \mathcal{R}_{i3} + Z_6 \mathcal{R}_{i1} \Big) C_0[ m_{G^\pm}^2, m_Z^2, m_{H_i}^2], \notag \\
  &G^{(C8)} _{+}= 0, \notag \\
  &F^{(C9)} _{+}= \frac{2s_W^2 m_W^2}{16 \pi^2 c_W} \mathcal{R}_{i1} \Big(\frac{1}{2}(Z_4 + Z_5) \mathcal{R}_{i2} - \frac{i}{2} (Z_4 - Z_5) \mathcal{R}_{i3} + Z_6 \mathcal{R}_{i1} \Big) C_0[ m_{H_i}^2, m_{W}^2, m_{G^\pm}^2], \notag \\
  &G^{(C9)} _{+}= 0, \notag \\
  &F^{(C10)}_{+} = \frac{2}{16 \pi^2 c_W} \frac{s_W^2 m_W^2}{v^2} \mathcal{R}_{i1} (\mathcal{R}_{i2} - i\mathcal{R}_{i3}) C_{24}[ m_{H_i}^2, m_{G^\pm}^2, m_W^2], \notag \\
  &G^{(C10)}_{+} = \frac{2m_W^2}{16 \pi^2 c_W} \frac{s_W^2 m_W^2}{v^2} \mathcal{R}_{i1} (\mathcal{R}_{i2} - i\mathcal{R}_{i3}) (C_{23} - C_{12})[ m_{H_i}^2, m_{G^\pm}^2, m_W^2], \notag \\
  &F^{(C11)}_{+} = \frac{2}{16 \pi^2 c_W} \frac{c_W^2 m_W^2}{v^2}  \mathcal{R}_{i1} (\mathcal{R}_{i2} - i\mathcal{R}_{i3}) \Big((D-1) C_{24} +(m_Z^2-m_W^2) C_0 \notag \\
  &\qquad \qquad -2 k_3 \cdot (k_2 C_{11} + k_3 C_{12}) + m_W^2 C_{21} + m_Z^2 C_{22} + 2 k_2 \cdot k_3 C_{23}  \Big)[ m_{H_i}^2, m_W^2, m_W^2], \notag \\
  &G^{(C11)}_{+} = \frac{2 m_W^2}{16 \pi^2 c_W} \frac{c_W^2 m_W^2}{v^2}  \mathcal{R}_{i1} (\mathcal{R}_{i2} - i\mathcal{R}_{i3}) (4C_{11} - 3C_{12} - C_{23})[ m_{H_i}^2, m_W^2, m_W^2], \notag \\
  &F^{(C12)}_{+} = \frac{-2}{16 \pi^2 c_W} \frac{m_W^2}{v^2} \mathcal{R}_{i1} (\mathcal{R}_{i2} - i\mathcal{R}_{i3}) \Big((D-1) C_{24} +4 (k_2 + k_3)\cdot k_2 C_0 + 2(2 k_2 + k_3)\cdot (k_2 C_{11} + k_3 C_{12}) \notag \\
  &\qquad \qquad + m_W^2 C_{21} + m_Z^2 C_{22} + 2 k_2 \cdot k_3 C_{23}  \Big)[ m_W^2, m_Z^2, m_{H_i}^2], \notag \\
  &G^{(C12)}_{+} = \frac{-2m_W^2}{16 \pi^2 c_W} \frac{m_W^2}{v^2} \mathcal{R}_{i1} (\mathcal{R}_{i2} - i\mathcal{R}_{i3}) \Big( -2C_0 + 2C_{11} -5C_{12} -C_{23}  \Big)[ m_W^2, m_Z^2, m_{H_i}^2], \notag \\
  &H^{(C~\mathrm{type})}_+ = 0.
\end{align}
where $c_W \equiv \cos\theta_W$, $s_W \equiv \sin \theta_W$, $c_{2W} \equiv \cos 2 \theta_W$ and $D$ is the space-time dimension.
In these formulae, we have omitted the summation of the indices for the neutral scalar bosons, and we have used a shorthand notation for the $C_{ij}$ functions as 
\begin{align}
  C_{ij}[m_a^2,m_b^2,m_c^2] \equiv C_{ij}[-k_2, -k_3; m_a^2,m_b^2,m_c^2].
\end{align}
The divergent parts of the $C$ type contributions are
\begin{align}
  &F_+^{(C1+C2)}|_{\mathrm{div}} = -\frac{s_W^2}{2} Z_7 \Delta, \notag \\
  &F_+^{(C4+C5)}|_{\mathrm{div}} = -\frac{s_W^2}{2} Z_6 \Delta, \notag \\
  &F_+^{(\mathrm{others}~C\mathrm{s})}|_{\mathrm{div}} = 0,
  \label{eq:Cdiv}
\end{align}
where we define the divergent quantity $\Delta$ as 
\begin{align}
  \Delta \equiv \frac{2}{16\pi^2 c_W} B_0 |_{\mathrm{div}}.
\end{align} 

Second, the $B$ type contributions are 
\begin{align}
  &F_+^{(B1)} = \frac{2}{16 \pi^2 c_W} \frac{s_W^2}{2} (\mathcal{R}_{i2} - i\mathcal{R}_{i3}) (Z_3 \mathcal{R}_{i1} + Z_7^R \mathcal{R}_{i2} - Z_7^I \mathcal{R}_{i3}) B_0[-k_1; m_{H^\pm}^2, m_{H_i}^2], \notag \\
  &F_+^{(B2)} = \frac{2}{16 \pi^2 c_W} \frac{s_W^2}{2} \mathcal{R}_{i1} \Big(\frac{1}{2}(Z_4 + Z_5) \mathcal{R}_{i2} - \frac{i}{2} (Z_4 - Z_5) \mathcal{R}_{i3} + Z_6 \mathcal{R}_{i1} \Big) B_0[-k_1; m_{G^\pm}^2, m_{H_i}^2], \notag \\
  &F_+^{(B3)} = \frac{2}{16 \pi^2 c_W}  \frac{s_W^2 m_W^2}{v^2} \mathcal{R}_{i1} (\mathcal{R}_{i2} - i\mathcal{R}_{i3}) B_0[-k_2; m_W^2, m_{H_i}^2], \notag \\
  &F_+^{(B4)} = \frac{2}{16 \pi^2 c_W}  \frac{s_W^2 m_Z^2}{v^2} \mathcal{R}_{i1} (\mathcal{R}_{i2} - i\mathcal{R}_{i3}) B_0[-k_3; m_Z^2, m_{H_i}^2], \notag \\
  &F_+^{(B5)} = \frac{2}{16 \pi^2 c_W} \frac{m_W^4 s_W^2}{v^2} \frac{1}{m_{H^\pm}^2 - m_W^2} \mathcal{R}_{i1} (\mathcal{R}_{i2} - i\mathcal{R}_{i3}) (B_0 - B_1) [-k_1; m_{H^\pm}^2, m_W^2], \notag \\
  &F_+^{(B6)} = \frac{2}{16 \pi^2 c_W} \frac{m_W^2 s_W^2}{2}  \frac{1}{m_{H^\pm}^2 - m_W^2} (\mathcal{R}_{i2} - i\mathcal{R}_{i3}) (Z_3 \mathcal{R}_{i1} + Z_7^R \mathcal{R}_{i2} - Z_7^I \mathcal{R}_{i3}) (B_0 +2 B_1) [-k_1; m_{H_i}^2, m_{H^\pm}^2], \notag \\
  &F_+^{(B7)} = \frac{2}{16 \pi^2 c_W} \frac{m_W^2 s_W^2}{2}  \frac{1}{m_{H^\pm}^2 - m_W^2} \mathcal{R}_{i1} \notag \\
  &\qquad \qquad \times \Big(\frac{1}{2}(Z_4 + Z_5) \mathcal{R}_{i2} - \frac{i}{2} (Z_4 - Z_5) \mathcal{R}_{i3} + Z_6 \mathcal{R}_{i1} \Big) (B_0 +2 B_1) [-k_1; m_{H_i}^2, m_{G^\pm}^2], \notag \\
  &F_+^{(B8)} = \frac{2}{16 \pi^2 c_W} \frac{m_W^2}{v^2} \frac{s_W^2}{2} \frac{-1}{m_{H^\pm}^2 - m_{G^\pm}^2} \mathcal{R}_{i1} (\mathcal{R}_{i2} - i \mathcal{R}_{i3}) \big(m_{H^\pm}^2 (B_0 - 2B_1) +\tilde{B_0} \big) [-k_1; m_{H_i}^2, m_W^2], \notag \\
  &F_+^{(B9)} = \frac{2}{16 \pi^2 c_W} \frac{s_W^2}{2} \frac{1}{m_{H^\pm}^2 - m_{G^\pm}^2} \Big(\frac{1}{2}(Z_4 + Z_5) \mathcal{R}_{i2} - \frac{i}{2} (Z_4 - Z_5) \mathcal{R}_{i3} + Z_6 \mathcal{R}_{i1} \Big) \notag \\
  &\qquad \qquad \times (Z_3 \mathcal{R}_{i1} + Z_7^R \mathcal{R}_{i2} - Z_7^I \mathcal{R}_{i3}) v^2 B_0[-k_1; m_{H_i}^2, m_{H^\pm}^2], \notag \\
  &F_+^{(B10)} = \frac{2}{16 \pi^2 c_W} \frac{s_W^2}{2} \frac{1}{m_{H^\pm}^2 - m_{G^\pm}^2} \Big(\frac{1}{2}(Z_4 + Z_5) \mathcal{R}_{i2} - \frac{i}{2} (Z_4 - Z_5) \mathcal{R}_{i3} + Z_6 \mathcal{R}_{i1} \Big) \notag \\
  &\qquad \qquad \times (Z_1 \mathcal{R}_{i1} + Z_6^R \mathcal{R}_{i2} - Z_6^I \mathcal{R}_{i3}) v^2 B_0[-k_1; m_{H_i}^2, m_{G^\pm}^2], \notag \\
  &G_+^{(B~\mathrm{type})} = H_+^{(B~\mathrm{type})}= 0,
\end{align}
where $\tilde{B}_0$ is defined by~\cite{Abbas:2018pfp}
\begin{align}
  \tilde{B_0}[k; m_a^2, m_b^2] \equiv (k^2 B_{21} + D B_{22})[k; m_a^2, m_b^2] = A_0[m_b^2] +m_a^2 B_0[k; m_a^2, m_b^2].
\end{align}
The divergent parts of the $B$ type contributions are given by
\begin{align}
  &F_+^{(B1)}|_{\mathrm{div}} = \frac{s_W^2}{2} Z_7 \Delta, \notag \\
  &F_+^{(B2)}|_{\mathrm{div}} = \frac{s_W^2}{2} Z_6 \Delta, \notag \\
  &F_+^{(B8)}|_{\mathrm{div}} = \frac{s_W^2}{2} \frac{m_W^2}{v^2} \frac{-1}{m_{H^\pm}^2 - m_{G^\pm}^2} \mathcal{R}_{i1} (\mathcal{R}_{i2} - i \mathcal{R}_{i3}) m_{H_i}^2 \Delta, \notag \\
  &F_+^{(B9)}|_{\mathrm{div}} = \frac{s_W^2}{2} \frac{1}{m_{H^\pm}^2 - m_{G^\pm}^2} \Big( \frac{1}{2} (Z_4 + Z_5) Z_7^R + \frac{i}{2} (Z_4 -Z_5) Z_7^I + Z_3 Z_6 \Big)v^2 \Delta, \notag \\
  &F_+^{(B10)}|_{\mathrm{div}} = \frac{s_W^2}{2} \frac{1}{m_{H^\pm}^2 - m_{G^\pm}^2}  \Big( \frac{1}{2} (Z_4 + Z_5) Z_6^R + \frac{i}{2} (Z_4 -Z_5) Z_6^I + Z_1 Z_6 \Big)v^2 \Delta, \notag \\
  &F_+^{(\mathrm{other}~B\mathrm{s})}|_{\mathrm{div}} = 0.
  \label{eq:Bdiv}
\end{align}

Third, the fermion contributions are given by 
\begin{align}
  &F^{(F1)}_+ = \frac{2 \sqrt{2} N_c}{16 \pi^2 c_W v} (V^\dagger)_{lm} \Bigg[ \Gamma_{ml}^u m_{u_m} \Big\{ m_{d_l}^2 (g_V^d - g_A^d) C_0 - (g_V^d + g_A^d) \big(g_{\alpha \beta} C^{\alpha \beta} - 2C_{24} + (2k_2 +k_3)\cdot (k_2 C_{11} \notag \\ 
  &\quad + k_3 C_{12}) + k_2 \cdot (k_2 + k_3) C_0  \big) \Big\} +\Gamma_{ml}^d m_{d_l} \Big\{ - (g_V^d + g_A^d) \big(g_{\alpha \beta} C^{\alpha \beta} - 2C_{24} + k_2 \cdot (k_2 C_{11} + k_3 C_{12})  \big) \notag \\
  &\quad + (g_V^d - g_A^d) \big(g_{\alpha \beta} C^{\alpha \beta} +(k_2 + k_3) \cdot (k_2 C_{11} + k_3 C_{12}) \big) \Big\}  \Bigg] [m_{u_m}^2, m_{d_l}^2, m_{d_l}^2], \notag \\
  &\quad +\frac{2 \sqrt{2}}{16 \pi^2 c_W v} \Gamma_{ll}^e m_{e_l} \Big\{ - (g_V^e + g_A^e) \big(g_{\alpha \beta} C^{\alpha \beta} - 2C_{24} + k_2 \cdot (k_2 C_{11} + k_3 C_{12})  \big) \notag \\
  &\quad + (g_V^e - g_A^e) \big(g_{\alpha \beta} C^{\alpha \beta} +(k_2 + k_3) \cdot (k_2 C_{11} + k_3 C_{12}) \big) \Big\} [0, m_{e_l}^2, m_{e_l}^2], \notag \\
  &G_+^{(F1)} = \frac{2 \sqrt{2} m_W^2 N_c}{16 \pi^2 c_W v} (V^\dagger)_{lm} \Bigg[ \Gamma^u_{ml} m_{u_m} (g_V^d + g_A^d) (2C_{12} + 2C_{23} + C_{11} + C_0) \notag \\
  &\quad+ \Gamma^d_{ml} m_{d_l} \Big\{ (g_V^d + g_A^d) (C_{12} + 2C_{23}) + (g_V^d - g_A^d) (C_{12} - C_{11}) \Big\} \Bigg] [m_{u_m}^2, m_{d_l}^2, m_{d_l}^2] \notag \\
  &\quad+\frac{2 \sqrt{2} m_W^2}{16 \pi^2 c_W v} \Gamma^e_{ll} m_{e_l} \Big\{ (g_V^e + g_A^e) (C_{12} + 2C_{23}) + (g_V^e - g_A^e) (C_{12} - C_{11}) \Big\} [0, m_{e_l}^2, m_{e_l}^2], \notag \\
  &H_+^{(F1)} = i \frac{2 \sqrt{2} m_W^2 N_c}{16 \pi^2 c_W v} (V^\dagger)_{lm} \Bigg[ \Gamma^u_{ml} m_{u_m} (g_V^d + g_A^d) (C_{11} + C_0) \notag \\
  &\quad + \Gamma^d_{ml} m_{d_l} \Big\{ (g_V^d + g_A^d) C_{12} - (g_V^d - g_A^d) (C_{12} - C_{11}) \Big\} \Bigg][m_{u_m}^2, m_{d_l}^2, m_{d_l}^2] \notag \\
  &\quad +i \frac{2 \sqrt{2} m_W^2 }{16 \pi^2 c_W v}  \Gamma^e_{ll} m_{e_l} \Big\{ (g_V^e + g_A^e) C_{12} - (g_V^e - g_A^e) (C_{12} - C_{11}) \Big\} [0, m_{e_l}^2, m_{e_l}^2], \notag \\
  &F^{(F2)}_+ = \frac{2 \sqrt{2} N_c}{16 \pi^2 c_W v} (V^\dagger)_{lm} \Bigg[ \Gamma_{ml}^d m_{d_l} \Big\{ m_{u_m}^2 (g_V^u - g_A^u) C_0 - (g_V^u + g_A^u) \big(g_{\alpha \beta} C^{\alpha \beta} - 2C_{24} + (2k_2 +k_3)\cdot (k_2 C_{11} \notag \\ 
  &\quad + k_3 C_{12}) + k_2 \cdot (k_2 + k_3) C_0  \big) \Big\} +\Gamma_{ml}^u m_{u_m} \Big\{ - (g_V^u + g_A^u) \big(g_{\alpha \beta} C^{\alpha \beta} - 2C_{24} + k_2 \cdot (k_2 C_{11} + k_3 C_{12})  \big) \notag \\
  &\quad + (g_V^u - g_A^u) \big(g_{\alpha \beta} C^{\alpha \beta} +(k_2 + k_3) \cdot (k_2 C_{11} + k_3 C_{12}) \big) \Big\}  \Bigg] [m_{d_l}^2, m_{u_m}^2, m_{u_m}^2] \notag \\
  &\quad +\frac{2 \sqrt{2}}{16 \pi^2 c_W v} \Gamma_{ll}^e m_{e_l} \Big\{ - (g_V^\nu + g_A^\nu) \big(g_{\alpha \beta} C^{\alpha \beta} - 2C_{24} + (2k_2 +k_3)\cdot (k_2 C_{11} + k_3 C_{12}) \notag \\
  &\quad + k_2 \cdot (k_2 + k_3) C_0  \big) \Big\} [m_{e_l}^2, 0, 0], \notag \\
  &G_+^{(F2)} = \frac{2 \sqrt{2} m_W^2 N_c}{16 \pi^2 c_W v} (V^\dagger)_{lm} \Bigg[ \Gamma^d_{ml} m_{d_l} (g_V^u + g_A^u) (2C_{12} + 2C_{23} + C_{11} + C_0) \notag \\
  &\quad + \Gamma^u_{ml} m_{u_m} \Big\{ (g_V^u + g_A^u) (C_{12} + 2C_{23}) + (g_V^u - g_A^u) (C_{12} - C_{11}) \Big\} \Bigg] [m_{d_l}^2, m_{u_m}^2, m_{u_m}^2], \notag \\
  &\quad +\frac{2 \sqrt{2} m_W^2}{16 \pi^2 c_W v} \Gamma^e_{ll} m_{e_l} (g_V^\nu + g_A^\nu) (2C_{12} + 2C_{23} + C_{11} + C_0) [m_{e_l}^2, 0, 0], \notag \\
  &H_+^{(F2)} = i \frac{2 \sqrt{2} m_W^2 N_c}{16 \pi^2 c_W v} (V^\dagger)_{lm} \Bigg[ - \Gamma^d_{ml} m_{d_l} (g_V^u + g_A^u) (C_{11} + C_0) \notag \\
  &\quad - \Gamma^u_{ml} m_{u_m} \Big\{ (g_V^u + g_A^u) C_{12} - (g_V^u - g_A^u) (C_{12} - C_{11}) \Big\} \Bigg][m_{d_l}^2, m_{u_m}^2, m_{u_m}^2] \notag \\
  &\quad -i \frac{2 \sqrt{2} m_W^2 }{16 \pi^2 c_W v} \Gamma^e_{ll} m_{e_l} (g_V^\nu + g_A^\nu) (C_{11} + C_0) [m_{e_l}^2,0,0], \notag \\
  &F^{(F3)}_+ = \frac{2 \sqrt{2} c_W N_c}{16 \pi^2 v} \frac{m_W^2 -m_Z^2}{m_W^2 - m_{H^\pm}^2} (V^\dagger)_{lm} \Big( \Gamma_{ml}^u m_{u_m} B_1 + \Gamma_{ml}^d m_{d_l} (B_0 + B_1) \Big)[-k_1; m_{d_l}^2, m_{u_m}^2] \notag \\
  &\quad +\frac{2 \sqrt{2} c_W}{16 \pi^2 v} \frac{m_W^2 -m_Z^2}{m_W^2 - m_{H^\pm}^2} \Gamma_{ll}^e m_{e_l} (B_0 + B_1) [-k_1; m_{e_l}^2, 0], \notag \\
  &F^{(F4)}_+ = \frac{2 \sqrt{2} s_W^2 N_c}{16 \pi^2 c_W v} \frac{1}{m_W^2 - m_{H^\pm}^2} (V^\dagger)_{lm} \Big\{ ( -\Gamma_{ml}^u m_{u_m} m_{d_l}^2 + \Gamma_{ml}^d m_{u_m}^2 m_{d_l}) B_0 \notag \\
  &\quad + (-\Gamma_{ml}^d m_{d_l} + \Gamma_{ml}^u m_{u_m}) (\tilde{B}_0 + k_1^2 B_1) \Big\}[-k_1; m_{d_l}^2, m_{u_m}^2] \notag \\
  &\quad -\frac{2 \sqrt{2} s_W^2}{16 \pi^2 c_W v} \frac{1}{m_W^2 - m_{H^\pm}^2} \Gamma_{ll}^e m_{e_l} (\tilde{B}_0 + k_1^2 B_1) [-k_1; m_{e_l}^2, 0], \notag \\
  &G_+^{(F3, F4)} = H_+^{(F3, F4)} = 0,
\end{align}
where the summation for the flavor indices $l,m$ is implicit, and $C^{\alpha \beta}$ is the $C$ type tensor function which satisfies the relation~\cite{Dreiner:2023yus}
\begin{align}
  &g_{\alpha \beta} C^{\alpha \beta} [k_1, k_2; m_a^2, m_b^2, m_c^2] = k_1^2 C_{21} + k_2^2 C_{22} + 2 k_1 \cdot k_2 C_{23} + D C_{24} \notag \\
  &= B_0[k_2; m_b^2, m_c^2] + m_a^2 C_0 [k_1, k_2; m_a^2, m_b^2, m_c^2].
\end{align}
The definition of the coupling constants are given by
\begin{align}
  &\Gamma_{lm}^u = (\rho^{u \dagger} V)_{lm}, ~~~~ \Gamma_{lm}^d = -(V \rho^d)_{lm}, ~~~~ \Gamma_{lm}^e = -(\rho^e)_{lm} \notag \\
  &g_V^f = \frac{1}{2}T_3^f - Q_f s_W^2, ~~~~ g_A^f = \frac{1}{2} T_3^f,
\end{align}
where $f = u,d,e,\nu$, and $T_3^f$ and $Q_f$ are the $\mathrm{SU}(2)_L$ isospin and the electromagnetic charge for $f$, respectively.
The divergent parts of the fermion contributions are given by 
\begin{align}
    F_+^{(F~\mathrm{type})}|_{\mathrm{div}} &= \frac{\sqrt{2} s_W^2 N_c}{v} \frac{1}{m_W^2 - m_{H^\pm}^2} \Big( (\rho^{u \dagger})_{ll} m_{u_l}^3 + (\rho^d)_{ll}m_{d_l}^3 \Big)\Delta  \notag \\
    &+ \frac{\sqrt{2} s_W^2}{v} \frac{1}{m_W^2 - m_{H^\pm}^2} (\rho^e)_{ll} m_{e_l}^3 \Delta ,
    \label{eq:Fdiv}
\end{align}

Finally, the $A$ type contributions are given by 
\begin{align}
  &F_+^{(A1)} = \frac{2}{c_W} \frac{s_W^2}{2} (\mathcal{R}_{i2} - i\mathcal{R}_{i3}) \frac{-1}{m_{H_i}^2} \frac{T_i}{v}, \notag \\
  &F_+^{(A2)} = \frac{2}{c_W} m_W^2 s_W^2 \frac{1}{2} (\mathcal{R}_{i2} - i\mathcal{R}_{i3}) \frac{1}{m_{H^\pm}^2 - m_W^2} \frac{-1}{m_{H_i}^2} \frac{T_i}{v}, \notag \\
  &F_+^{(A3)} = \frac{2}{c_W} \frac{s_W^2}{2} \Big(\frac{1}{2}(Z_4 + Z_5) \mathcal{R}_{i2} - \frac{i}{2} (Z_4 - Z_5) \mathcal{R}_{i3} + Z_6 \mathcal{R}_{i1} \Big) v^2 \frac{1}{m_{H^\pm}^2 - m_{G^\pm}^2} \frac{-1}{m_{H_i}^2} \frac{T_i}{v}, \notag \\
  &F_+^{(A4)} = \frac{2}{c_W} \frac{s_W^2}{2} \frac{1}{m_{H^\pm}^2 - m_{G^\pm}^2} \Pi_{H^+ G^-}, \notag \\
  &G_+^{(A~\mathrm{type})} = H_+^{(A~\mathrm{type})}= 0,
\end{align}
where we denote the tadpole and the self energy of $H^+ G^-$ as 
\begin{align}
  T_i \equiv 
  \begin{minipage}[b]{0.1\linewidth}
  \centering
  \setlength{\feynhandlinesize}{0.5pt}
  \setlength{\feynhandarrowsize}{4pt}
  \begin{tikzpicture}[baseline=-0.15cm]
  \begin{feynhand}
      \vertex[NWblob] (a) at (0,0) {};
      \vertex[dot] (b) at (0,-0.4) {};
      \vertex (b) at (0,-0.7) {$H_i$};
  \end{feynhand}
  \end{tikzpicture}
  \end{minipage}, ~~~~~~~
  \Pi_{H^+ G^-} \equiv 
  \begin{minipage}[b]{0.1\linewidth}
  \centering
  \setlength{\feynhandlinesize}{0.5pt}
  \setlength{\feynhandarrowsize}{4pt}
  \begin{tikzpicture}[baseline=-0.15cm]
      \begin{feynhand}
      \vertex (a) at (0,0){$H^+$}; \vertex (d) at (1.2,0); \vertex (g) at (1.2,0.8); \vertex (h) at (2.4,0) {$G^-$};
      \propag[chasca] (a) to (d);
      \propag[chasca] (d) to (h);
      \propag[sca] (d) to [half right, looseness=1.6] (g);
      \propag[sca] (g) to [half right, looseness=1.6] (d);
      \end{feynhand}
  \end{tikzpicture}
  \end{minipage}
  \qquad \qquad.
  \notag 
\end{align}
By using the relation between $\mathcal{R}$ and $\mathcal{M}^2$ in eq.~(\ref{eq:relation_neutmass_diagmass}), the $A$ type contributions can be simplified as 
\begin{align}
  F_+^{(A~\mathrm{type})} = \frac{2}{c_W} \frac{s_W^2}{2} \frac{1}{m_{H^\pm}^2 - m_W^2} \Big( - (\mathcal{R}_{i2} - i\mathcal{R}_{i3}) \frac{T_i}{v} + \Pi_{H^+ G^-} \Big).
\end{align}
When we write the tadpole of $H_i$ as $T_i^{(p)}$, where a particle $p$ runs inside the loop, they are given by
\begin{align}
  &T_i^{(G^\pm)} = \frac{1}{16 \pi^2 v} (Z_1 \mathcal{R}_{i1} + Z_6^R \mathcal{R}_{i2} - Z_6^I \mathcal{R}_{i3}) v^2 A_0[m_{G^\pm}^2], \notag \\
  &T_i^{(G^0)} = \frac{1}{16 \pi^2 v} \frac{1}{2} (Z_1 \mathcal{R}_{i1} + Z_6^R \mathcal{R}_{i2} - Z_6^I \mathcal{R}_{i3}) v^2 A_0[m_{G^0}^2], \notag \\
  &T_i^{(H^\pm)} = \frac{1}{16 \pi^2 v} (Z_3 \mathcal{R}_{i1} + Z_7^R \mathcal{R}_{i2} - Z_7^I \mathcal{R}_{i3}) v^2 A_0[m_{H^\pm}^2], \notag \\
  &T_i^{( H_j )} = \frac{1}{16 \pi^2 v}  (\lambda_{ijj}^H +\lambda_{jij}^H +\lambda_{jji}^H ) v^2 A_0[m_{H_j}^2], \notag \\
  &T_i^{(u_l)} = -\frac{\sqrt{2}N_c}{16 \pi^2} \Big( \mathcal{R}_{i2} (\rho_u + \rho_u^\dagger )_{ll} - i\mathcal{R}_{i3} (\rho_u - \rho_u^\dagger )_{ll} \Big) m_{u_l} A_0[m_{u_l}^2], \notag \\
  &T_i^{(d_l)} = -\frac{\sqrt{2}N_c}{16 \pi^2} \Big( \mathcal{R}_{i2} (\rho_d + \rho_d^\dagger )_{ll} + i\mathcal{R}_{i3} (\rho_d - \rho_d^\dagger )_{ll} \Big) m_{d_l} A_0[m_{d_l}^2], \notag \\
  &T_i^{(e_l)} = -\frac{\sqrt{2}}{16 \pi^2} \Big( \mathcal{R}_{i2} (\rho_e + \rho_e^\dagger )_{ll} + i\mathcal{R}_{i3} (\rho_e - \rho_e^\dagger )_{ll} \Big) m_{e_l} A_0[m_{e_l}^2].
\end{align}
We here have defined $\lambda_{ijk}^H$ as 
\begin{align}
  \lambda_{ijk}^H &\equiv \frac{1}{2} \mathcal{R}_{i1} \mathcal{R}_{j1} \big( Z_1 \mathcal{R}_{k1} + 3Z_6^R \mathcal{R}_{k2} - 3 Z_6^I \mathcal{R}_{k3} \big) \notag \\ 
  &+ \frac{1}{2} \mathcal{R}_{j2} \mathcal{R}_{k2} \big( (Z_3 + Z_4 + Z_5) \mathcal{R}_{i1} + Z_7^R \mathcal{R}_{i2} - Z_7^I \mathcal{R}_{i3} \big) \notag \\
  &+ \frac{1}{2} \mathcal{R}_{j3} \mathcal{R}_{k3} \big( (Z_3 + Z_4 - Z_5) \mathcal{R}_{i1} + Z_7^R \mathcal{R}_{i2} - Z_7^I \mathcal{R}_{i3} \big).
\end{align}
We note that the gauge and ghost fields cause the tadpole diagram for $H_i$, however, they do not give the $A$ type contributions.
This is because the scalar-gauge-gauge and the scalar-ghost-ghost interactions are proportional to $\mathcal{R}_{i1}$, and the $A$ type contributions from these tadpole are proportional to $\sum_i \mathcal{R}_{i1} (\mathcal{R}_{i2} - i\mathcal{R}_{i3}) = 0$.
The self energies $\Pi_{H^+ G^-}^{(p)}$, where a scalar particle $p$ runs inside the loop, are given by 
\begin{align}
  &\Pi_{H^+ G^-}^{(G^\pm)} = \frac{1}{16\pi^2 v^2} 2 Z_6 v^2 A_0[m_{G^\pm}^2], \notag \\
  &\Pi_{H^+ G^-}^{(G^0)} = \frac{1}{16\pi^2 v^2} \frac{1}{2} Z_6 v^2 A_0[m_{G^0}^2], \notag \\
  &\Pi_{H^+ G^-}^{(H^\pm)} = \frac{1}{16\pi^2 v^2} 2 Z_7v^2 A_0[m_{H^\pm}^2], \notag \\
  &\Pi_{H^+ G^-}^{(H_i)} = \frac{1}{16\pi^2 v^2} \frac{1}{2} \Big( (Z_4 + Z_5)\mathcal{R}_{i1}\mathcal{R}_{i2} -i (Z_4 - Z_5)\mathcal{R}_{i1}\mathcal{R}_{i3} \notag \\
  &\qquad \qquad + Z_6 \mathcal{R}_{i1}\mathcal{R}_{i1} + Z_7 (\mathcal{R}_{i2}\mathcal{R}_{i2}+\mathcal{R}_{i3}\mathcal{R}_{i3}) \Big) v^2 A_0[m_{H_i}^2].
\end{align}
The divergent parts of the $A$ type contributions are given by 
\begin{align}
  &F^{(A~\mathrm{type})}_+ |_{\mathrm{div}} = \frac{s_W^2}{2} \frac{1}{m_{H^\pm}^2 - m_W^2} (Z_6 m_W^2 + Z_7 m_{H^\pm}^2) \Delta \notag \\
  &+\frac{s_W^2}{2} \frac{1}{m_{H^\pm}^2 - m_W^2} \frac{v^2}{2} \Big( -2Z_7 \frac{m_{H^\pm}^2}{v^2} -2Z_1 Z_6 -2 Z_3 Z_6 - Z_4 Z_6 - Z_4 Z_7 - Z_5 Z_6^* - Z_5 Z_7^* \Big) \Delta \notag \\
  &-\frac{\sqrt{2} s_W^2}{v} \frac{1}{m_W^2 - m_{H^\pm}^2 } \Big( N_c (\rho_u^\dagger)_{ll} m_{u_l}^3 + N_c (\rho_d)_{ll} m_{d_l}^3  + (\rho_e)_{ll} m_{e_l}^3 \Big) \Delta.
  \label{eq:Adiv}
\end{align}
From eqs.~(\ref{eq:Cdiv}), (\ref{eq:Bdiv}), (\ref{eq:Fdiv}) and (\ref{eq:Adiv}), all of the divergences are cancel out.

By using the above results, we can derive the coefficients of the tensor decomposition for the decay $H^- \to W^- Z$.
The coefficients $F_-$, $G_-$ and $H_-$ from the $A$, $B$ and $C$ type contributions can be obtained by just taking complex conjugate of the coupling constants in front of the loop functions in the above results.
The fermion contributions of $F_-$, $G_-$ and $H_-$ can be obtained by replacing as 
\begin{align}
  \Gamma^f_{ml} \to \big(\Gamma^{f \dagger} \big)_{lm}~~~~(f = u,d,e),~~~~(V^\dagger)_{lm} \to V_{ml},
\end{align}
and $i \to -i$ in $F_+^{(F~\mathrm{type})}$, $G_+^{(F~\mathrm{type})}$ and $H_+^{(F~\mathrm{type})}$.

\bibliographystyle{unsrt}
\bibliography{refs}

\end{document}